\documentclass[preprintnumbers,amsmath,amssymb,nofootinbib]{revtex4}
\usepackage{hyperref}
\usepackage{subfigure}
\usepackage{amssymb}
\usepackage{graphicx}
\usepackage{amsmath}
\usepackage{color}
\usepackage{multirow}

\newcommand{\mgFull}{\texttt{MadGraph5\_aMC@NLO v.2.7.2}}
\definecolor{pantoneCB}{rgb}{0.0588235, 0.298039, 0.505882}
\hypersetup{bookmarks=true,unicode=true,pdftoolbar=true,pdfmenubar=true,
	pdffitwindow=false,
	pdfnewwindow=true,colorlinks=true,allcolors=pantoneCB}
\newcommand{\nn}{\nonumber}
\newcommand{\lsim}{\mathrel{\mathop{\kern 0pt \rlap
			{\raise.2ex\hbox{$<$}}}
		\lower.9ex\hbox{\kern-.190em $\sim$}}}
\newcommand{\gsim}{\mathrel{\mathop{\kern 0pt \rlap
			{\raise.2ex\hbox{$>$}}}
		\lower.9ex\hbox{\kern-.190em $\sim$}}}
\newcommand{\be}{\begin{equation}}
\newcommand{\ee}{\end{equation}}
\newcommand{\bea}{\begin{eqnarray}}
\newcommand{\eea}{\end{eqnarray}}

\begin{document}

\preprint{IITH-PH-0007/20}

\title{Obscure Higgs boson at colliders}

\author{Priyotosh Bandyopadhyay}
\email{bpriyo@phy.iith.ac.in}
\affiliation{Indian Institute of Technology Hyderabad, Kandi,  Sangareddy-502287, Telengana, India}

\author{Antonio Costantini}
\email{antonio.costantini@bo.infn.it}
\affiliation{INFN - Sezione di Bologna, via Irnerio 46, 40126 Bologna, Italy}

\begin{abstract}In this study we consider an extension of the Standard Model with a complex hypercharge zero triplet scalar. In this scenario one of the charged Higgs bosons remains purely triplet and does not couple to the fermions, making it elusive at colliders. Also the physical pseudoscalar is a pure triplet and this purity makes it a suitable dark matter candidate without the need of discrete symmetries, unlike other extensions. The bounds from relic density and direct dark matter search experiments select its mass to be $\sim 1.35-1.60$  TeV. The pure triplet charged Higgs gives rise to displaced signatures and their sensitivity at LHC and MATHUSLA have been studied. The prospects at present and future hadron/muon colliders of such exotic scalars are  pointed out by calculating their productions cross-section and dominant decay modes. We present also the expected reach for the triplet states at a multi-TeV muon collider.
\end{abstract}


\maketitle

\section{Introduction}

The Higgs boson discovery was the last key stone of the Standard Model (SM) \cite{Aad:2012tfa,Chatrchyan:2012ufa}. However, the SM Higgs boson mass aces the quadratic divergence as it is not protected by any symmetry like chiral symmetry or gauge symmetry. Supersymmetry came as astounding solution in canceling the quadratic divergence  and in its minimal extension  with $R$-parity provides the much needed dark matter candidate.  The Minimal Supersymmetric Standard Model (MSSM) predicts four physical Higgs bosons: one charged Higgs, two CP-even and one CP-odd. The lightest CP-even Higgs boson mass in the MSSM is predicted to be lower than the $Z$ boson mass, at tree-level. On the contrary,  the observed Higgs boson mass is around $125.5$ GeV: this demands large loop-corrections with either a heavy SUSY (supersymmetry) mass scale or a highly fine-tuned parameter space \cite{Carena:2011aa}. An extension of the Higgs sector softens the amount of quantum corrections needed contributing to the Higgs mass both at tree- as well as loop-level. This makes SUSY at the TeV scale a theoretical reality \cite{Ellwanger:2009dp,Bandyopadhyay:2013lca,Bandyopadhyay:2014tha,Bandyopadhyay:2015oga, Bandyopadhyay:2015tva}. 

However, the non-observation of any other beyond-the-Standard-Model (BSM) states situated these Higgs bosons masses to rather high values or somehow are not probed so far. The intriguing quest one might ask is that if there are Higgs bosons in the mass ranges already probed by LHC but still not visible. One naive possibility is if the Higgs boson that is produced potentially decays in invisible modes. The other 
striking possibility is that if such scalars are feebly produced at hadron colliders. Such theoretical possibility arise when the Higgs bosons coupling to fermions is highly suppressed, which inhibits both the productions via quarks/gluons and the decay channels in the fermionic modes. This kind of scenario will provides the guideline of this paper. Because the SM does not have a right-handed $SU(2)$ doublet, an $SU(2)$ triplet Higgs boson with zero hypercharge  cannot couples directly with the fermions. A Higgs boson in such representation, together with an extra $Z_2$ symmetry can provide the much needed dark matter \cite{Ayazi:2015mva,Khan:2016sxm,Jangid:2020qgo}. In a non-supersymmetric (non-SUSY) framework we can extend the SM with a real $Y=0$ triplet, conversely to the corresponding supersymmetric case, where we need to have a complex $Y=0$ triplet along with two Higgs doublets to fulfill the anomaly free condition and holomorphicity of superpotential \cite{Bandyopadhyay:2013lca,Bandyopadhyay:2014tha,Bandyopadhyay:2014vma, Bandyopadhyay:2015oga,Bandyopadhyay:2015tva,Bandyopadhyay:2015ifm,Bandyopadhyay:2017klv}. Hence it's clear that the minimal supersymmetric and non-supersymmetric extensions of the SM are rather different. Specifically, in the real case we do not have any pseudoscalar and the charged Higgs bosons are conjugate to each other \cite{Ayazi:2015mva,Khan:2016sxm,Chabab:2018ert,Jangid:2020qgo}. The choice of a complex $Y=0$ representation in a non-SUSY framework certainly invokes an extra physical pseudoscalar. However such a pseudoscalar will have no room to mix with the doublet Higgs, unlike other SUSY/non-SUSY extensions, e.g. the MSSM or the Two-Higgs-Doublet-Model (2HDM) \cite{Coleppa:2019cul}. This purity fabricates the pseudoscalar as dark matter candidate because, without adding any discrete symmetry, its cubic interactions with fermions and gauge bosons cease to exist. In a nutshell, adding a $Y=0$ complex triplet to the Higgs sector of the SM brings a natural dark matter candidate, the physical pseudoscalar.

This article is organized as follows. In Section~\ref{sec:model} we briefly discuss the main features of the model, the electroweak symmetry breaking  along with the CP-conserving mass eigenstates of the Higgs sector  and its custodial limit. The phenomenology of the model is examined in Section \ref{sec:pheno}: we address its dark matter content, the main physics of long-lived BSM particles and the effect of the extra-scalars on the trilinear and quartic Higgs self-couplings. In Section \ref{sec:collider} we present the results for the main collider signatures of the model at present and future facilities. We draw our conclusions in Section~\ref{sec:concl}.

\section{Complex triplet Extension of the Standard Model}\label{sec:model}

In this Section we discuss the extension of the Standard Model with a complex triplet with $Y=0$, which we name complex Triplet extension of the Standard Model (cTSM). The gauge and fermion sectors are identical to the Standard Model ones, and we do not write them here. The only difference with the SM lays in the scalar sector, where apart from the usual Higgs doublet
\bea
\Phi=\left(
\begin{array}{c}
\phi^+\\
\Phi_0
\end{array}
\right),
\eea
we consider a complex triplet with $Y=0$ hypercharge, namely
\bea
T=\frac{1}{\sqrt2}\left(
\begin{array}{cc}
t_0&\sqrt2\, t_1^+\\
\sqrt2\, t_2^-&-t_0
\end{array}
\right).
\eea
We stress that, as a consequence of being a complex multiplet, $(t_1^+)^*\neq t_2^-$ and $t_0$ is also complex. The neutral component of $H$ and $T$ will acquire a vacuum expectation value (VEV) and break the electroweak symmetry,
\bea
\Phi_0&=& \frac{1}{\sqrt2}\left(v+\phi_0 + i\, \sigma_0\right),\\
t_0&=& \frac{1}{\sqrt2}\left(v_T+\phi^t_0 + i\, \sigma^t_0\right).
\eea
After the spontaneous electroweak symmetry breaking (EWSB) the scalars mix and gauge bosons and fermions became massive via the Brout-Engler-Higgs mechanism \cite{Englert:1964et,Higgs:1964pj,Guralnik:1964eu}. In particular, the masses of the gauge bosons are given by
\begin{equation}
m_W=\frac{1}{2}g_2 \sqrt{v^2+4 v_T^2}\;,\quad m_Z=\frac{1}{2}\sqrt{\left(g_1^2+g_2^2\right)}v , \label{eq:mwz}
\end{equation}
where $g_1$ and $g_2$ are the gauge coupling constant of the $U(1)_Y$ and $SU(2)_L$ groups respectively.
It is well known that such an extension of the SM will not respect the custodial symmetry, manifested by the fact that 
\bea
\rho=\frac{m_W^2}{m_Z^2 \cos^2\theta_w}\neq1
\eea
at tree-level. The experimental value of the $\rho$-parameter is \cite{Zyla:2020zbs}
\bea
\rho^{\texttt{ex}}=1.00038\pm0.00020,
\eea 
and this will constraint the allowed values of $v_T \lesssim 5$ GeV .

\subsection{Scalar potential and mass matrices}\label{sec:CP}
Although not explicitly stated, we have assumed that we are in a situation where CP-symmetry is not violated spontaneously ($v>0,v_T>0$) nor explicitly. The explicit CP violation occur if one consider complex parameters for the scalar potential. In this paper we will consider only the scenario where CP-symmetry is not violated. In the CP-conserving case the potential of the model is
\begin{equation}\label{vpot}
V=V_{1}+V_{2},
\end{equation}
with
\bea\label{cpcpot}
V_{1}&=&\mu^2 \Phi^\dagger \Phi +\frac{\lambda_H}{2}\Phi^\dagger \Phi \Phi^\dagger \Phi + m^2_T\, \texttt{tr}[T^\dagger T] + \frac{\lambda_T}{2} \texttt{tr}[T^\dagger T]\texttt{tr} [T^\dagger T] + \frac{\lambda_{T'}}{2}\texttt{tr}[T^\dagger T\, T^\dagger T]\nn\\
&&\quad+\frac{\lambda_{HT}}{2}\Phi^\dagger \Phi \,\texttt{tr}[T^\dagger T]
+\kappa_{HT}\, (\texttt{tr}[\Phi^\dagger T \Phi] + \texttt{h.c.}),
\eea
and
\bea\label{extrapot}
V_{2}&=&\Big(m'^2_T\, \texttt{tr}[T\, T] + \frac{\lambda_T^{(2)}}{2} \texttt{tr}[T\, T\, T\, T]+ \frac{\lambda_T^{(3)}}{2} \texttt{tr}[T^\dagger T\, T\, T]\nn\\
&&\qquad+\frac{\lambda_{HT}^{(2)}}{2}\Phi^\dagger \Phi \,\texttt{tr}[T\, T]
\Big) + \texttt{h.c.}.
\eea
Here $\mu, m_T$ and $m'_T$ are SM Higgs bosons and the complex triplet mass terms respectively, $\lambda_H, \, \lambda_{T,T'},\,\lambda_{HT},\,\lambda_{T,HT}^{(i)}$ and $\kappa_{HT}$ are dimensionless and dimensionful couplings for scalars respectively. As already stated, the parameters entering in Eq.~\ref{vpot} are assumed to be real. Let us notice that in writing Eq.~\ref{extrapot} we make use of the trace relations 
\begin{equation}
\mathtt{tr}[T\,T\,T\,T]=\frac{1}{2} (\mathtt{tr}[T\,T])^2\;,\; \mathtt{tr}[T^\dagger T\,T\,T]= \frac{1}{2}\mathtt{tr}[T^\dagger T] \mathtt{tr}[T\,T] .
\end{equation}
The gauge-fixing part of the lagrangian is
\begin{equation}
\mathcal{L}_{gf}= - \frac{1}{2 \xi_V} \left( \partial^\mu V_\mu + \xi_V M_V G_V
\right)^2
\end{equation}
in the case of a massive gauge boson $V_\mu$, whose Goldstone boson is $G_V$.

After EWSB the scalars mix, and the conditions for the minimum of the potential are given by
\bea
\mu^2&=&-\frac{\lambda_H\, v^2}{2} +\kappa_{HT}v_T-\frac{\lambda_{HT}}{4}v_T^2-\frac{\lambda_{HT}^{(2)}}{2}v_T^2,\\
m_T^2&=&-\frac{\lambda_T\, v_T^2}{2}-\frac{\lambda_{T'}\, v_T^2}{4}+\kappa_{HT}\frac{v^2}{2v_T}-\frac{\lambda_{HT}v^2}{4}-2 m'^2_T-\frac{\lambda_{HT}^{(2)}}{2}v^2 - (\lambda_T^{(2)}+\lambda_T^{(3)})v_T^2 .
\eea
We define the mixing in the scalar sector as
\bea\label{gaugees}
h_i=\mathcal{R}^S_{ij}\,H_j\;,\quad a_i=\mathcal{R}^P_{ij}\,A_j\;,\quad h_i^+=\mathcal{R}^C_{ij}\,H^+_j ,
\eea
where $\vec{H}=(\phi_0,\,\phi_0^t)$, $\vec{A}=(\sigma_0,\,\sigma_0^t)$, $\vec{H}^+=(\phi^+,\,(t_2^-)^*,\,t_1^+)$ and $\mathcal{R}^{S,P,C}$ are the rotation matrices for scalar, pseudoscalar and charged Higgs bosons respectively. The mass matrices for  CP-even, CP-odd neutral scalar and charged scalars are given below,
\bea\label{mms}
\mathcal{M}^S=\left(
\begin{array}{cc}
 \lambda_{H} v^2 &(\lambda_{HT}+2\lambda_{HT}^{(2)})\frac{v\, v_T}{2}-\kappa_{HT}v \\
\cdot & \frac{1}{2v_T}(\kappa_{HT}v^2+(2\lambda_T+\lambda_{T'}+2(\lambda_T^{(2)}+\lambda_T^{(3)})) v_T^3)\\
\end{array}
\right),
\eea

\bea\label{mmp}
\mathcal{M}^P=\left(
\begin{array}{cc}
 \frac{1}{4} v^2 \,\xi_Z (g_2 \cos \theta_w+g_1 \sin \theta_w)^2 & 0
   \\
 \cdot & \kappa_{HT}\frac{v^2}{2v_T}-4m'^2_T-\lambda_{HT}^{(2)}v^2 - (4\lambda_T^{(2)}+\lambda_T^{(3)})v_T^2\\
\end{array}
\right),\nn\\
\eea

\bea\label{mmc}
\mathcal{M}^C=\left(
\begin{array}{ccc}
 \frac{1}{4} g_2^2 \xi_W v^2+2 \kappa_{HT}\, v_T & \frac{v}{2\sqrt2}(2\kappa_{HT} -g_2^2 \xi_W v_T)& \frac{v}{2\sqrt2}(2\kappa_{HT}  -g_2^2 \xi_W v_T) \\
\cdot& \frac{\kappa_{HT}\,v^2}{2v_T}+\frac{v_T^2}{2} g_2^2\xi_W - \tilde m &\frac{v_T^2}{2}g_2^2\xi_W + \tilde m \\
 \cdot& \cdot &\frac{\kappa_{HT}\,v^2}{2v_T}+\frac{v_T^2}{2} g_2^2\xi_W - \tilde m\\
\end{array}
\right),\nn\\
\eea
with $\tilde m = 2m'^2_T+\lambda_{HT}^{(2)}/2\,v^2 + (\lambda_T^{(2)}+\lambda_T^{(3)}/2-\lambda_{T'}/4)v_T^2$. For our analysis we have chosen the unitary gauge, where $\xi_Z=\xi_W\equiv0$. Looking at Eq.~\ref{mmp} we can conclude that the physical pseudoscalar of the model will be a pure triplet state even after EWSB.

\subsection{Eigenvalues and Eigenvectors of the Scalar Sector}
A remarkable feature of the cTSM is that we can write analytical expressions for both eigenvalues and eigenvectors of the scalar spectrum. The pseudoscalar sector is by far the simplest. The mass of the physical pseudoscalar is given by
\bea\label{mps}
m^2_{a_P}= \kappa_{HT}\frac{v^2}{2v_T}-4m'^2_T-\lambda_{HT}^{(2)}v^2 - (4\lambda_T^{(2)}+\lambda_T^{(3)})v_T^2,
\eea
and the other pseudoscalar is the neutral Goldstone boson. The structure of this physical pseudoscalar boson in terms of the gauge eigenstates is given by
\bea\label{cpcpsstr}
a_P=\sigma_0^t .
\eea
Thus the physical pseudoscalar is a pure triplet, the reason being that it's orthogonal to the neutral Goldstone, which is $a_0\equiv G_Z= \sigma_0$  and  hence the nomenclature $a_P$. This feature has some important consequences. The most important is that $a_P$ does not couple with the fermions, nor at tree-lever or at loop order. The physical pseudoscalar neither has the tree-level cubic interactions with gauge bosons nor with $h_i\,Z$. We remind that the vertex $A\, H_i Z$ is non-zero in the 2HDM as well as in supersymmetric scenarios \cite{Bandyopadhyay:2014vma,Jangid:2020qgo}. The absence of a coupling with the fermions means that it will have no loop-level couplings with the vector bosons as well. This rare quality promotes the triplet pseudoscalar to be a candidate dark matter, if it is the lightest among the other triplets.

Similar to the pseudoscalar sector, even the physical charged Higgses have a pure state. After EWSB, the expressions of the charged Higgs bosons in terms of their gauge-eigenstates is
\bea
	h_T^+&=&\frac{2 v_T}{\sqrt{v^2+4v_T^2}}\,\phi^+ +\frac{2v}{\sqrt2\sqrt{v^2+4v_T^2}}\,(t_2^-)^* + \frac{2v}{\sqrt2\sqrt{v^2+4v_T^2}}\,t_1^+\label{cpccht},\\
	h_P^+&=&-\frac{1}{\sqrt2}\,(t_2^-)^* + \frac{1}{\sqrt2}\, t_1^+\label{cpcchp},\\
h_0^+&=&-\frac{v}{\sqrt{v^2+4v_T^2}}\, \phi^+ + \frac{\sqrt2 v_T}{\sqrt{v^2+4v_T^2}}\, (t_2^-)^* + \frac{\sqrt2 v_T}{\sqrt{v^2+4v_T^2}}\,t_1^+ \label{chgold}.
\eea
Here $h_0^\pm\equiv G_W$ is the charged Goldstone boson that exhibit a mixing between doublet and triplet degrees of freedom. The same is true for the mostly-triplet charged Higgs $h_T^+$. The triplet part of the Goldstone boson is complemented by the doublet part of $h_T^+$. Conversely $h_P^+$ remains a pure state even after EWSB.

As we can see from Eq.~\ref{chgold} the rotation angles $\mathcal{R}^C_{0i}$ of the charged Goldstone are functions only of the VEVs of the neutral scalars. The Goldstones are in fact the fingerprint of EWSB mechanism, which will take place when the neutral scalars develop  VEVs and hence their expression cannot be affected by other parameters of the potential. Even in this case the nomenclature chosen for the massive charged Higgs bosons is related to their structure in terms of the gauge eigenstates. The masses for these two physical charged Higgs bosons are given by
\bea
m^2_{h_T^\pm}&=&\kappa_{HT}\left(\frac{v^2}{2\,v_T}+2\,v_T\right)\label{mcht},\\
m^2_{h_P^\pm}&=& \kappa_{HT}\frac{v^2}{2\,v_T}-4m'^2_T-\lambda_{HT}^{(2)}v^2 - (2\lambda_T^{(2)}+\lambda_T^{(3)}+\frac{\lambda_{T'}}{2})v_T^2.\label{mchp}
\eea
Comparing Eq.~\ref{mchp} and Eq.~\ref{mps} we can see that
\begin{equation} \label{massdiff}
m^2_{h_P^\pm}-m^2_{a_P}= \left(2\lambda_{T}^{(2)}+\frac{\lambda_{T'}}{2}\right)v_T^2
\end{equation}
at tree-level.

Unlike the pseudoscalar and the charged Higgs sectors, in the CP-even neutral sector both the CP-even neutral scalars are mixed states of doublet and triplets. Their expression, in the limit $v_T\ll v$, is given in Eq.~\ref{h1str} and Eq.~\ref{h2str} respectively,
\bea
	h_D&=&\frac{1}{N_{h_D}}\Big((8\,v^2\kappa_{HT}^3+\ldots)\,\phi_0+16\,\kappa_{HT}^3v\,v_T\,\phi_0^t\Big),\label{h1str}\\
	h_T&=&\frac{1}{N_{h_T}}\left((-2\kappa_{HT}v_T+(\lambda_{HT}+2\lambda_{HT}^{(2)}-4\lambda_H)v_T^2)\,\phi_0 +\kappa_{HT} \,v\, \phi_0^t\right)\label{h2str}.
\eea
Here $N_{h_D/T}$ are normalization factors. Looking at Eq.~\ref{h1str}, we can see that the coefficient of $\phi_0$, which is the doublet contribution, is $\sim v^2+\mathcal O(v_T^k)$, whereas the triplet part is $\sim v\, v_T$. Hence the neutral scalar $h_D$ is a mostly-doublet state, and compatible with the observed Higgs boson around 125.5 GeV. The opposite is true for $h_T$, which is a triplet-like state. At the order $\mathcal{O}(v_T^2)$ the masses of these CP-even scalars are given by
\bea
m^2_{h_D}&=&\lambda_H v^2-2\kappa_{HT}\,v_T+2\left(\lambda_{HT}+2\lambda_{HT}^{(2)}-2\lambda_H\right)v_T^2\label{msd},\\
m^2_{h_T}&=&\frac{\kappa_{HT}}{2v_T}\left(v^2+4v_T^2\right)+\left(4\lambda_{H}-2\lambda_{HT}-4\lambda_{HT}^{(2)}+\lambda_T+\frac{\lambda_{T'}{2}}+2(\lambda_T^{(2)}+\lambda_T^{(3)})\right)v_T^2,\label{mst}
\eea
where the model parameters get constraints from the recent Higgs boson mass and branching measurements at ATLAS and CMS experiments at the LHC \cite{ATLAS:2018doi,Sirunyan:2018koj}, which will be discussed later.

Let us remark an essential features of the model. The coupling of the neutral scalars with the fermions are proportional to the coefficient of $\phi_0$ given in Eq.~\ref{h1str} and Eq.~\ref{h2str}. The reason is because the triplet does not have direct interaction with the fermions. The $\phi_0$ coefficient of $h_T$ is related to $v_T/v$ and this means that the coupling of $h_T$ to the fermions is highly suppressed. 

Let us also have a closer look on the vertices involving the pure states of the model $a_P$ and $h_P^\pm$ and the gauge bosons $W^\pm$, $Z$. Here we have the couplings $a_P\, h_i^+W^-$ and $Z\, h_i^+W^-$ in terms of the rotation angles, as defined earlier,
\bea
	g_{a_P\, h_{i=P,\, T}^+W^-}&=&-\frac{g_L}{2}\left(\mathcal R^P_{21}\mathcal R^C_{i1}-\sqrt2 \mathcal R^P_{22}(\mathcal R^C_{i2}-\mathcal R^C_{i3})\right),\label{purev}\\
	g_{Z\, h_{i=P,\, T}^+W^-}&=&-\frac{i}{2}g_L\left(g_Y v \sin\theta_W \mathcal R^C_{i1}+\sqrt2 g_L v_T\cos\theta_W (\mathcal R^C_{i2}-\mathcal R^C_{i3})\right).\label{purev1}
\eea
It is interesting to see that whenever we have only one (odd numbers) pure triplet state in the vertex, i.e. $a_P$ or $h_P^\pm$, these two couplings vanish. It can be seen using the explicit expressions of the rotations for the charged Higgs bosons in Eqs.~\ref{purev}-\ref{purev1}, i.e. $g_{a_P\, h_{T}^+W^-},\, g_{Z\, h_{P}^+W^-}$ vanish but $g_{Z\, h_{T}^+W^-}$ remains non-zero. However the appearance of such pure triplet states twice in a vertex (e.g. $a_P h_P^+ W^-$) makes it non-zero. The pure triplet nature acts effectively as an odd number in a discrete $Z_2$ symmetry.

\subsection{Discrete Symmetry Limit}
In this section we consider a limit where we use an additional discrete symmetry to reduce some of the terms in the potential Eq.~\ref{vpot}. If we apply a $Z_3$ symmetry such that the triplet transforms as $T \to e^{\frac{2\pi i}{3}}T $ whereas the remaining particles are unaffected by the symmetry, then all the terms in Eq.~\ref{extrapot} will vanish.  At the EWSB scale when triplet also breaks the custodial symmetry it can create domain wall problem, due to the spontaneous breaking of the discrete symmetry \cite{Vilenkin:1984ib,Ellwanger:2009dp}. To prohibit such scenario we break explicitly the discrete symmetry with the soft term $\kappa_{HT}$ in Eq.~\ref{cpcpot}. In this limit the pure charged Higgs and the pure pseudoscalar bosons become nearly degenerate leaving $h_T^+$ slightly heavier, by a factor of $2\,\kappa_{HT}\, v_T$, as shown in Eq.~\ref{mtL}, 
\begin{equation}
m^2_{a_P}=\kappa_{HT}\frac{v^2}{2\,v_T},\quad m^2_{h_P^\pm}=\kappa_{HT}\frac{v^2}{2\,v_T}+\frac{\lambda_{T'}}{2}v_T^2,\quad m^2_{h_T^\pm}=\kappa_{HT}\left(\frac{v^2}{2\,v_T}+2\,v_T\right)\label{mtL}.
\end{equation}

We can see from Eq.~\ref{mtL}  that, in the limit $\lambda_{T'}\to0$, the pure states and other charged Higgs boson masses are proportional to $\kappa_{HT}$ and are restricted by the choice of of $v_T$ as well. In this scenario, the splitting between the pure charged state and pseudoscalar can come from the quantum corrections and it is around $166$ MeV \cite{Cirelli:2009uv}\footnote{It has been shown that the large quartic couplings can cause the loosing of perterbative unitarity of the model \cite{Jangid:2020qgo}. We check that for the cTSM by calculating the beta functions at one-loop for the dimensionless couplings, which are given in Appendix \ref{sec:rge}. For the choices of our benchmark points and for the enhanced perturbative validity of the dimensionless coupling, we will consider the  $\lambda_{T^\prime}\sim 0$ limit for the rest of the analysis.}.
It is interesting to note that the forms of the charged Higgs states Eq.~\ref{cpccht}-~\ref{chgold} are the same even in this limit. However the same does not occur for the CP-even scalar eigenstates, see Eq.~\ref{h1str}-~\ref{h2str}. The analytical expressions of these eigenvectors depends on the parameters of $V_2$. 

Given these considerations, for the phenomenological analysis of the cTSM we confine ourself to the case $V_2\equiv0$. The \textit{minimal set} of invariants included in $V_1$ encompass the relevant physics of this extension of the SM, viz. the presence of a pure triplet pseudoscalar as a dark matter candidate as well as the presence of a pure triplet charged Higgs. Moreover, in the phenomenological analysis we will consider only the scenario $\lambda_{T'}\to0$. In this limit the pure charged Higgs boson becomes a long-lived particle and can be searched with dedicated experiment like MATHUSLA \cite{Lubatti:2019vkf}, as discussed later.  A non-zero $\lambda_{T'}$, however can only splits the degeneracy  by $\frac{\lambda_{T^\prime}}{2}v^2_T \lesssim 1$ GeV, given that both $v_T$ and $\lambda_{T'}$ are restricted, resulting in a reduction of the decay length of the pure charged Higgs boson.

\subsection{Custodial-Symmetric Limit}\label{cust}
Before concluding this section we briefly comment on the scalar spectrum in the inert-triplet scenario.
This case corresponds to the restoration of the custodial symmetry at tree-level. From Eq.~\ref{eq:mwz} it is clear that in the limit $v_T\to0$ we obtain $\rho\equiv1$ at tree-level. In this case the scalar spectrum is further simplified w.r.t. the case $v_T\neq0$. In fact we have a degeneration among the triplet states and all states are pure, i.e. either doublet or triplet. The massive scalar spectrum is then given by
\bea\label{cusmass}
m^2_{h_D}&=& \lambda_H v^2,\quad h_D=\phi_0,\\
m^2_{h_T}&=& m_T^2 + 2m'^2_T + \frac{1}{2}(\frac{\lambda_{HT}}{2}+\lambda_{HT}^{(2)})v^2,\quad h_T=\phi^t_0,\\
m^2_{a_P}&=&m_T^2 - 2m'^2_T + \frac{1}{2}(\frac{\lambda_{HT}}{2}-\lambda_{HT}^{(2)})v^2,\quad a_P=\sigma_0^t,\\
m^2_{h_{T/P}^+}&=&m_T^2 \pm 2m'^2_T + \frac{1}{2}(\frac{\lambda_{HT}}{2}\pm\lambda_{HT}^{(2)})v^2,\quad h^+_{T/P}=\frac{1}{\sqrt2}(t_1^+\pm(t_2^-)^*).
\eea
The limit $v_T\to0$ outlines a peculiar scenario. Apart from the doublet neutral scalar, which is now the SM Higgs boson, we have two pairs of degenerate states that do not couple with the fermions at tree level. They can be generated in pairs from the SM Higgs boson or the massive gauge bosons. Such scenario resembles the inert real triplet one \cite{Khan:2016sxm,Jangid:2020qgo} with the possibility of either $h_T$ or $a_T$ being a dark matter candidate.

\section{Phenomenology of the cTSM}\label{sec:pheno}

In this section we address the relevant phenomenology of the cTSM, focusing on three different topics, the presence of a Dark Matter (DM) candidate, the possibility of long-lived (LL) heavy states and the self-interaction of the Higgs boson(s). As we are going to see, the first two possibilities are closely related to each other.

\subsection{Dark Matter}

The possibility of a scalar DM candidate has attracted considerable attention in the recent years. The Weakly-Interactive Massive Particle (WIMP) paradigm was first proposed in the context of the MSSM, where the natural DM candidate is a fermion \cite{Jungman:1995df}.  However there is no need for SUSY if one want to address the problem of DM. The possibility of minimalistic models with a scalar or fermionic DM candidate has then been analyzed from a more general point of view \cite{Cirelli:2005uq,  Bandyopadhyay:2017bgh, Bandyopadhyay:2017tlq, Jangid:2020qgo,Han:2020uak}. Even in the case of triplet extensions of the SM constraint were imposed on the mass of the stable particle that acts as DM state. There are some difference of course giving the detail of the model considered, but for a scalar DM the representation of the gauge group plays a crucial role.

In spite of the various analysis that can be found in literature, it seems that not every possibility has been considered. In fact, in the context of the cTSM, we naturally have a DM candidate in the massive pseudoscalar $a_P$. As we have explained in the previous section, the massive pseudoscalar is a pure triplet state. Thus it has no coupling with the fermions and hence no coupling with the photons. Its pseudoscalar nature forbids the coupling with the massive gauge bosons. In other context, such as the extension of the SM with a real triplet, these features are a consequence of an imposed $Z_2$ parity, which assign parity $+1$ to the SM fields and parity $-1$ to the triplet scalar \cite{Khan:2016sxm,Jangid:2020qgo,FileviezPerez:2008bj,Chiang:2020rcv}. In this way the neutral component of the triplet multiplet became the DM candidate. Apart from the triplet extensions of the SM, models with extra scalars in smaller representations of $SU(3)_c\times SU(2)_L\times U(1)_Y$ were also considered. Even in these scenarios extra symmetries (global and/or local) are needed to ensure the stability of the DM candidate \cite{LopezHonorez:2006gr,Dolle:2009fn,Goudelis:2013uca,Arcadi:2018pfo,Camargo:2019ukv,Bandyopadhyay:2017tlq}. 

In the cTSM, instead of imposing a $Z_2$ parity (or enlarging the gauge group), we have a \textit{purity} symmetry for the pseudoscalar which behaves similarly. Of course this is a special feature of the massive pseudoscalar and it is related to the orthogonality between $a_P$ and $G_Z$, the Goldstone of the $Z$ boson. We would emphasize that such a symmetry is not imposed on the Lagrangian but naturally emerges as a consequence of the interplay between the matter content and the gauge symmetry of the model. 

The mass of the scalar DM candidate for which the observed relic density is correctly achieved lays in the TeV range \cite{Cirelli:2005uq,Cirelli:2009uv,Jangid:2020qgo}.

\subsection{Parameter scan}\label{param}
In order to obtain the correct relic density for the DM candidate, we have implemented the model in \texttt{MadDM v.3.0} \cite{Ambrogi:2018jqj} with the help of \texttt{SARAH-4.14.1} \cite{Staub:2013tta} for the generation of the Universal FeynRules Output (UFO). The scalar spectrum has been obtained through a scan over the parameter space with
\bea\label{scan1}
v_T\in[0,5]\,\textrm{GeV},\quad \lambda_{H,\,HT,\,T} \in[-3/2,3/2],\quad \kappa_{HT}\in[10,500]\,\textrm{GeV}.
\eea
We have selected the benchmark points (BPs) with
\bea \label{scan2}
m_{h_D}=125.18\pm0.16\,\textrm{GeV},\quad \big|\mathcal{R}^S_{11}\big|>99/100 .
\eea

\begin{figure}[t!]
	\centering
	\includegraphics[scale=.65]{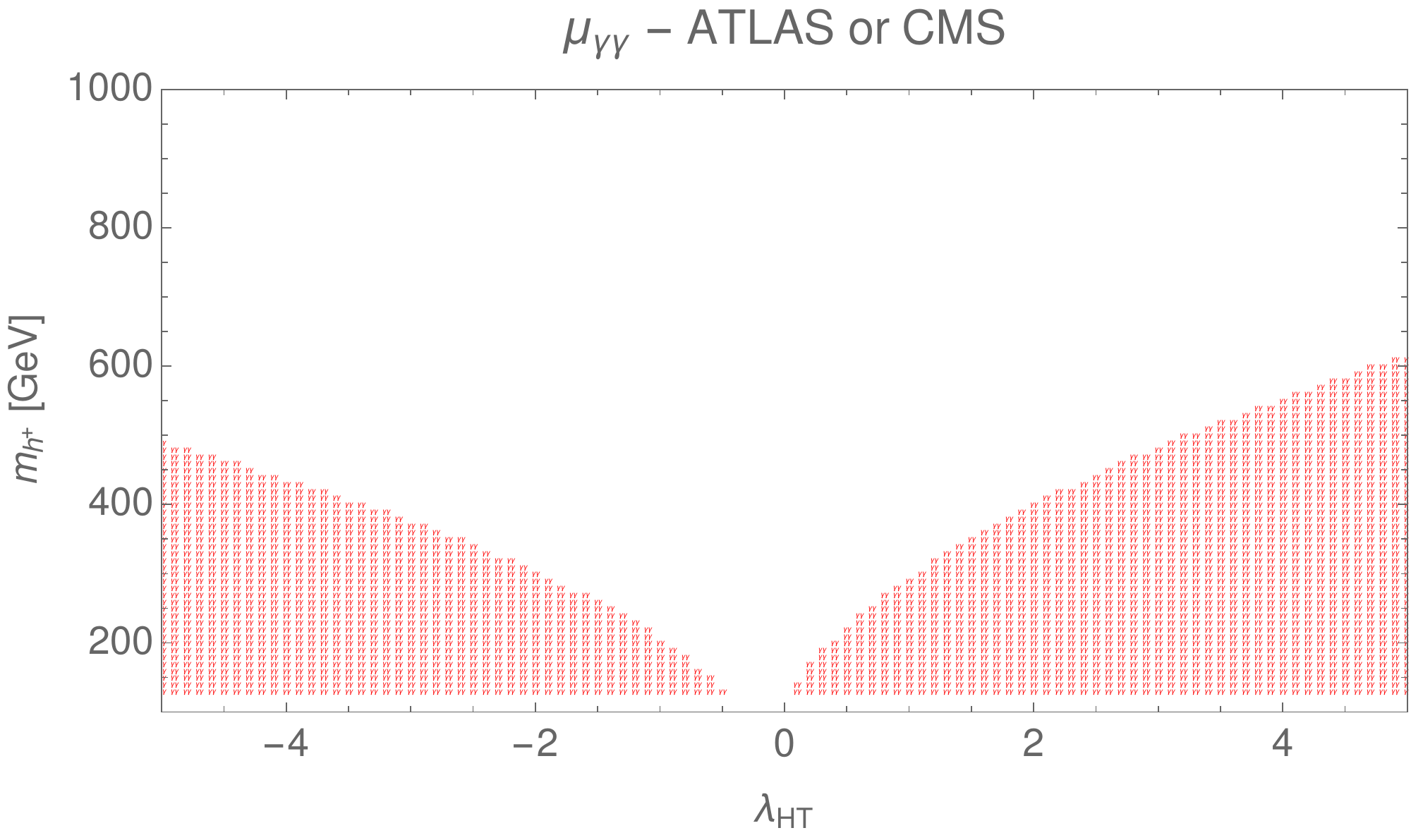}
\caption{Charged Higgs boson mass as a function of $\lambda_{HT}$. The region in red is excluded by the recent measurements of the diphoton signal strength of ATLAS or CMS \cite{Aaboud:2018xdt,Sirunyan:2018ouh}. Here $h^\pm \equiv h_{P/T}^\pm$ and their masses are considered to be the same.}\label{mudiphoF}
\end{figure} 
The second condition implies that $h_D$ (considered as the lightest neutral scalar) has SM-like couplings with quarks, leptons and massive gauge bosons, satisfying the recent bounds from LHC \cite{ATLAS:2018doi,Sirunyan:2018koj}. This also implies that the coupling of $h_D$ with two gluons has the correct SM-value. The situation is slightly different for the coupling of $h_D$ with two photons. In fact in the cTSM we have two massive charged scalars that enters in the loop-induced interaction $h_D\,\gamma\,\gamma$. The partial decay width of $h_D$ in diphotons is given by \cite{Djouadi:2005gj}
\bea
\Gamma_{h_D\to\gamma\gamma}&=&\frac{G_\mu \alpha^2 m_{h_D}^3}{128\sqrt2\pi^3}\Big|\sum_f N_c\, Q_f g_{h_D ff}A_{1/2}^h(\tau_f)+g_{h_D VV}A^h_1(\tau_W)\\
&&\quad+\sum_s\frac{m^2_W}{2\cos\theta_W^2 m^2_{h_s^\pm}}g_{h_D h_s^\pm h_s^\mp}A^h_0(\tau_s)\Big|^2.\nn
\eea
Here $A_{0,1/2,1}^\Phi(\tau)$ are the scalar, fermion and vector one-loop functions respectively \cite{Djouadi:2005gi}. The coupling $g_{h_D h_s^\pm h_s^\mp}$ is the trilinear interaction of the lightest Higgs with the charged Higgs bosons, normalized to $i\, m_Z^2/v$. For the two charged Higgs bosons of the cTSM we have
\bea
g_{h_D h_P^\pm h_P^\mp} &\equiv&\frac{\lambda_{h_D h_P^\pm h_P^\mp}}{i\, m_Z^2/v}=\frac{1}{i\, m_Z^2/v}(-i)\big(\lambda_{HT}\,v\,\mathcal{R}^S_{11}+(2\lambda_T+3\lambda_{T'}) v_T \mathcal{R}^S_{12}\big),\\
g_{h_D h_T^\pm h_T^\mp} &\equiv&\frac{\lambda_{h_D h_T^\pm h_T^\mp}}{i\, m_Z^2/v}\simeq\frac{1}{i\, m_Z^2/v}(-i)\left(\lambda_{HT}\,v\,\mathcal{R}^S_{11}+8\kappa_{HT} \frac{v_T}{v}\mathcal{R}^S_{11}+(2\lambda_T+\lambda_{T'}) v_T\mathcal{R}^S_{12}\right).
\eea
In light of the recent results for the Higgs in diphoton signal strength \cite{Aaboud:2018xdt, Sirunyan:2018ouh}, defined as $\mu_{\gamma\gamma}=\Gamma_{h\to\gamma\gamma}^{\textrm{SM}}/\Gamma_{\Phi\to\gamma\gamma}$, we have to consider BPs compatible with
\bea\label{diphosign}
\mu^{\textrm{ATLAS}}_{\gamma\gamma}=0.99^{+0.15}_{-0.14}\;,\quad \mu^{\textrm{CMS}}_{\gamma\gamma}=1.10^{+0.20}_{-0.18}.
\eea
In Figure~\ref{mudiphoF} we present the allowed regions of  $m_{h_{T/P}^\pm}-\lambda_{HT}$  from the diphoton signal strength in the cTSM
\footnote{We set $\kappa_{HT}=500$ GeV and $v_T=5$ GeV. The exclusion region has a mild dependence on $\kappa_{HT}$ and $v_T$.}. 
The masses of both $h_P^\pm$ and $h_T^\pm$ are considered to be the same
\footnote{Their small difference ($\lsim\mathcal{O}(1)$ GeV) wouldn't affect much our results for the diphoton constraint.}
, cfr. Eqs~(\ref{mcht}) and (\ref{mchp}). The region in red is excluded by the recent results of the Higgs in diphoton signal strength \cite{Aaboud:2018xdt, Sirunyan:2018ouh}. We can see that a charged Higgs boson with $m_{h^\pm}\gsim 600$ GeV is compatible with the recent LHC data.

Next we analyze the constraints coming from the dark matter analysis, i.e. from DM relic calculation and direct DM searches. In this scenario the pure triplet pseudoscalar $a_P$ is the DM and due to its $SU(2)$ charge it dominantly annihilates to $W^+ W^-$. Its annihilation to $ZZ$ is less dominant. There is also a co-annihilation channel via $a_P h_P^\pm \to Z W^\pm$ whereas $a_P h_P^\pm \to t b/ c s$ are the sub-dominant ones. Moreover because the purity acts like a discrete symmetry the co-annihilation cross-section of $a_P h_T^\pm \to Z W^\pm$ is zero.

The parameter space has been scanned as shown in Eq.~\ref{scan1} and Eq.~\ref{scan2}. We've used \texttt{MadDM v.3.0} \cite{Ambrogi:2018jqj} to compute the relic abundance of the DM candidate. The measured value of this important cosmological parameter is \cite{Ade:2015xua}
\bea\label{obsrel}
(\Omega_{DM}\,h^2)_{\texttt{exp.}}=0.1198\pm0.0015 .
\eea


\begin{figure}[t!]
	\centering
	\mbox{\subfigure[]{\includegraphics[scale=.35]{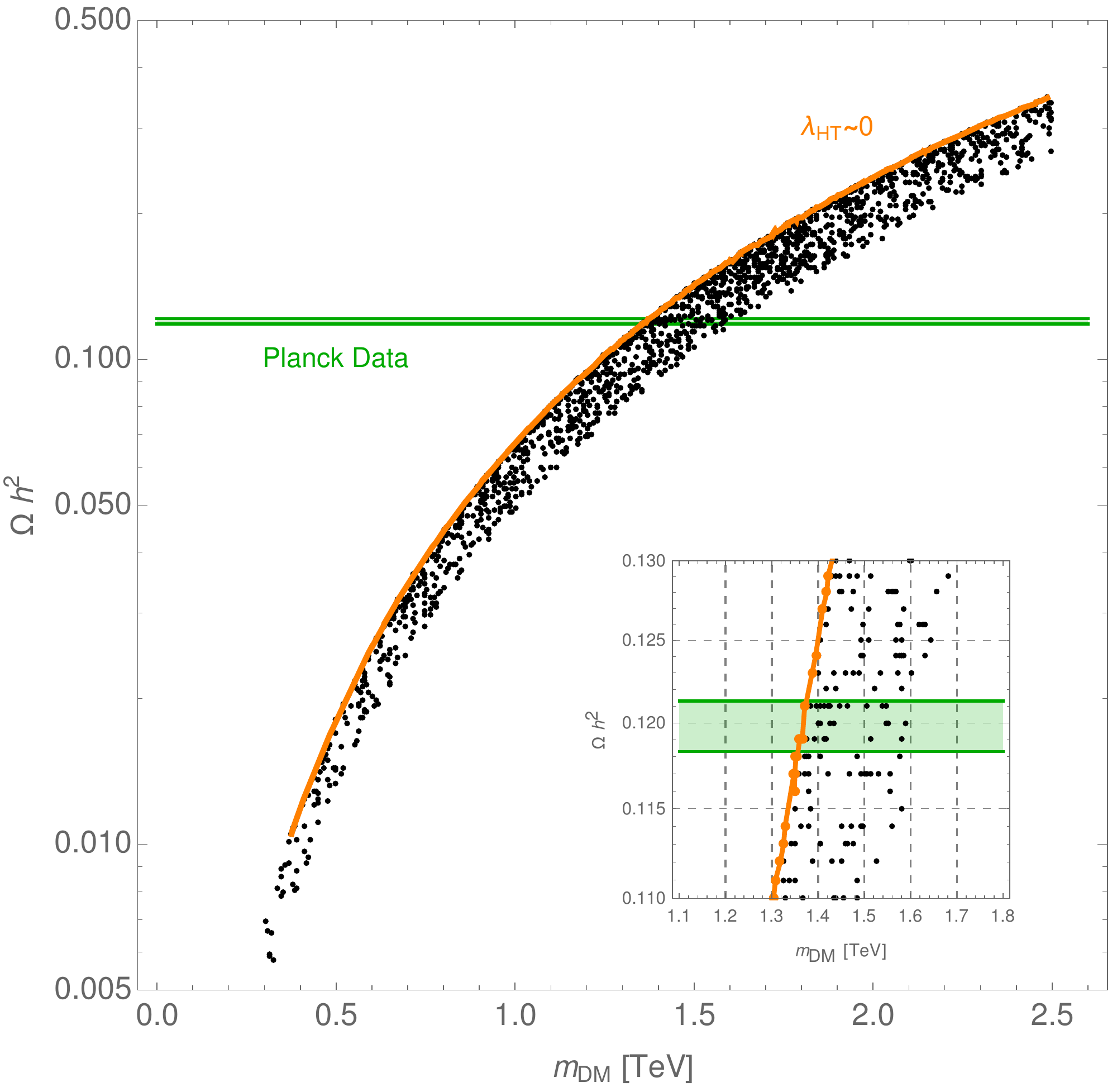}}\hspace{.5cm}\subfigure[]{\includegraphics[scale=.35]{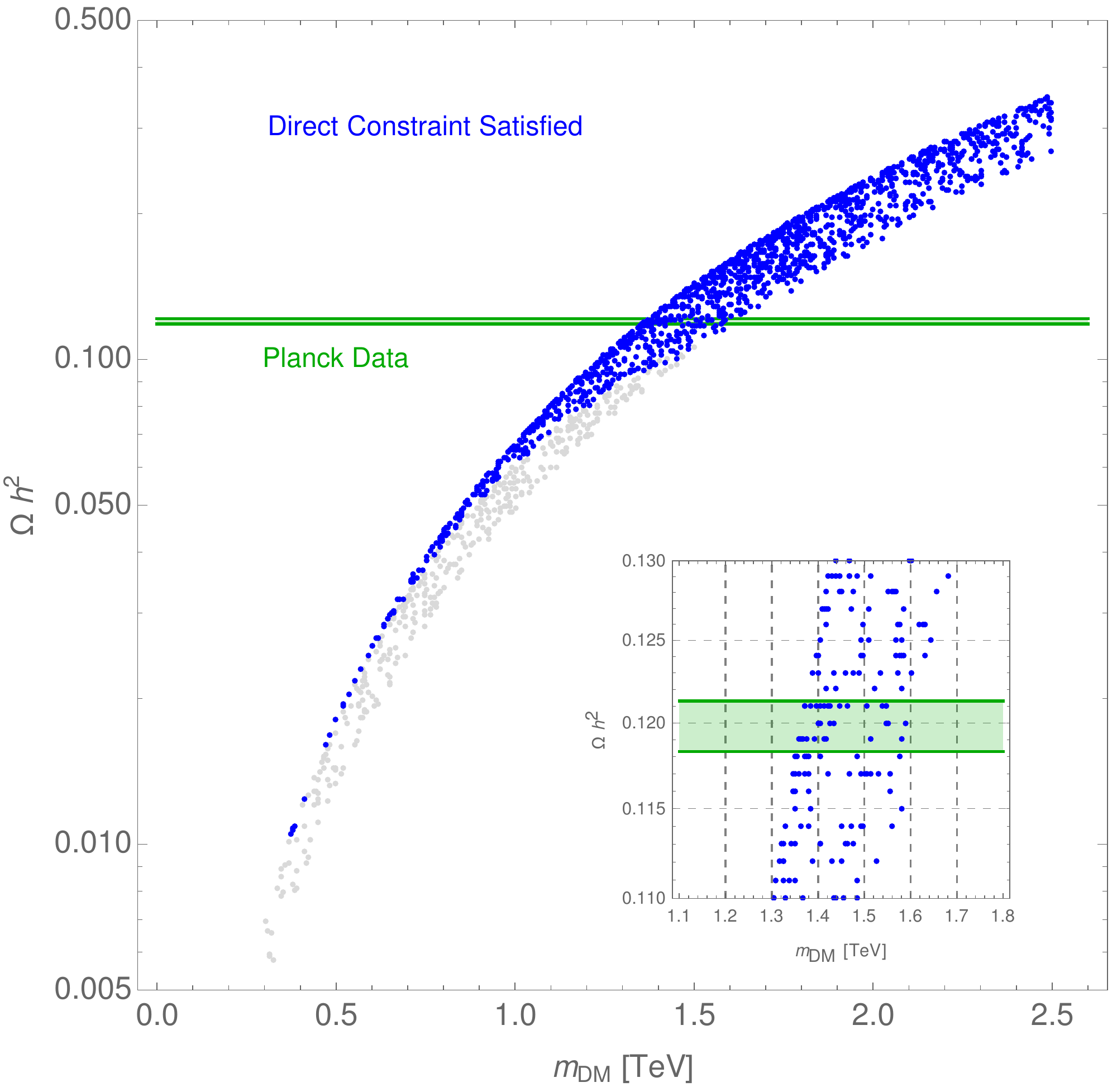}}
	}
	\mbox{\subfigure[]{\includegraphics[scale=.6]{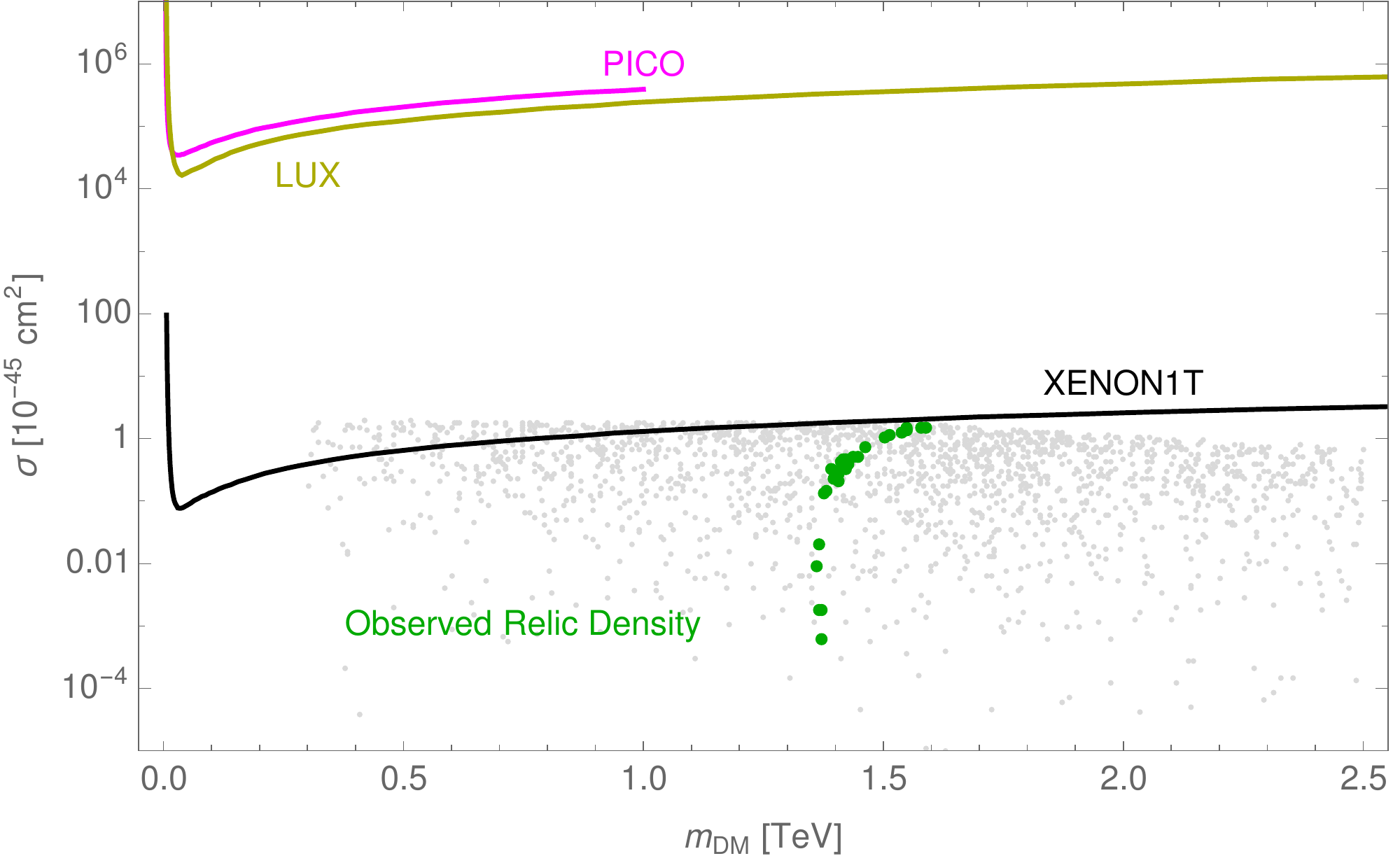}}}
	\caption{(a) Relic density as a function of the DM mass. The black points satisfy the LHC constraints on Higgs couplings. We have marked in orange the points with $\lambda_{HT}=0$. The green area represent the Planck results \cite{Ade:2015xua}. (b) Relic density as a function of the DM mass. In blue the points that are allowed by the direct searches \cite{Aprile:2017iyp,Amole:2017dex,Akerib:2017kat}. (c) DM-N cross-section as a function of the DM mass. In black, magenta and yellow we plot the constraints coming from XENON1T \cite{Aprile:2017iyp} , PICO \cite{Amole:2017dex} and LUX \cite{Akerib:2017kat} respectively. We mark in green the points that satisfy the Planck constraint on relic abundance.}\label{relicF}
\end{figure} 
\begin{figure}[t!]
	\centering
	\includegraphics[scale=1]{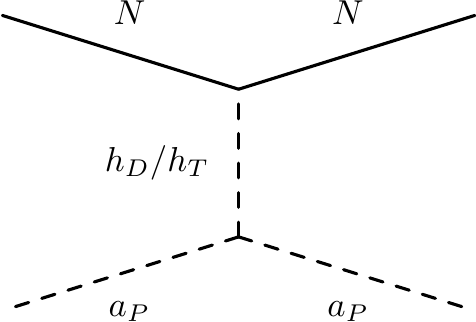}
	\caption{Dark matter - nucleon scattering in the cTSM. The process is mediated by the neutral scalars $h_D$ and $h_T$. Their interaction with the fermions is proportional to $\mathcal{R}^S_{11}$ and $\mathcal{R}^S_{21}$ respectively.}\label{fig:dmn}
\end{figure} 

We present our results in Figure~\ref{relicF}. In particular, Figure~\ref{relicF}(a) shows $\Omega h^2$ versus $m_{DM}$. The black points satisfy Eq.~\ref{scan2} together with the constraint from the diphoton signal, given in Eq.~\ref{diphosign}. We've enlightened in orange the points with $\lambda_{HT}=0$ and the green band represent the Planck constraint on the relic abundance, cfr. Eq~\ref{obsrel}. By looking at the zoomed plots of Figure \ref{relicF}(a), we can see that there is a minimum value for the DM mass for which the correct DM relic can be obtained and which is around 
\bea
m^{min}_{DM}\equiv m^{min}_{a_P}\sim1.35\,\textrm{TeV},
\eea
and corresponds to the $\lambda_{HT}\sim0$ case. The maximum possible value of the DM mass,  consistent with the observed relic density for the scan, is
\bea
m^{max}_{DM}\equiv m^{max}_{a_P}\sim1.60\,\textrm{TeV}.
\eea

Next we consider the direct dark matter detection bounds where the DM candidate must fulfill the constraint on the DM-nucleon (DM-N) cross-section. The DM-N scattering can take place as shown in  Figure~\ref{fig:dmn}. We remind that the pure pseudoscalar $a_P$ cannot couples directly with fermions, however its interactions to the neutral triplet- or doublet-like scalars $h_T/h_D$ make a way out for the DM-nucleon scattering. Such  interaction with quarks is proportional to its doublet component, as explained before. 

The  tree-level DM-nucleon scattering cross-section  is given by
\begin{equation}\label{DMsc}
	\sigma^{tree}_{DM-N}\approx\frac{4}{3\pi}s_{\alpha_T}^2c_{\alpha_T}^2\frac{m_N^2\mu^6_{DM-N}}{m_{DM}^2v^2 v_T^2}\frac{(m_{h_D}^2-m_{h_T}^2)^2}{m_{h_D}^4m_{h_T}^4} v_{DM}^4,
\end{equation}
where we denote $s_{\alpha_T}=\sin\alpha_T$ as the sine of the mixing angle between $h_D$ and $h_T$, the mediators of the $DM\;N \to DM\;N$ scattering  \cite{Azevedo:2018exj,Arina:2019tib}. After imposing the constraints  in Figure \ref{relicF}(b) depicted by blue points, we see there are plenty of points allowed by both DM relic and direct DM search constraints.

In Figure~\ref{relicF}(b)-(c) we present the direct detection constraint on the DM candidate. Specifically, Figure~\ref{relicF}(b) is the correlation plot between $\Omega h^2$ and $m_{DM}$ where the blue point are allowed by the direct DM searches \cite{Aprile:2017iyp,Amole:2017dex,Akerib:2017kat}.  We plot in Figure~\ref{relicF}(c) the cross-section versus DM mass for our scanned data points. It is evident that most of the points are allowed by the bounds coming from different experiments measuring the spin-independent cross-section, like XENON1T \cite{Aprile:2017iyp} or the spin dependent cross-section, like PICO \cite{Amole:2017dex} and LUX \cite{Akerib:2017kat}. These bounds are shown in black, red and yellow respectively in the plot. The green points are those satisfying the correct relic density. 

Moreover, from Figure \ref{relicF}(b) we can conclude that if $\lambda_{HT}\sim0$ a pseudoscalar with $m_{a_P}<1.35$ TeV is still compatible with the direct detection constraint(s), although the relic density in this case is below the observed one. This might suggest the possibility of a DM sector, composed by a heavy BSM particle and other (unspecified) physical objects. 

In light of this result for the mass of the DM candidate of the cTSM we can reconsider the reason for $v_T$ being a small parameter, $v_T\sim\mathcal{O}(1)$ GeV. In fact, from the expression
\bea
m^2_{a_P}=\frac{\kappa_{HT}}{2}\frac{v^2}{v_T},
\eea
we obtain that
\bea
v_T=\frac{\kappa_{HT}}{2}\frac{v^2}{\Lambda_{DM}^2}\simeq \frac{\kappa_{HT}}{2}\times 0.03\, \textrm{GeV}.
\eea 
The cTSM has then only two scales, namely $\Lambda_{EW} \sim v$ and $\Lambda_{DM} \sim m_{DM}$.

\subsection{Benchmark points}
Having analyzed the DM content of the cTSM we are able to select points in the parameter space that satisfy the current constraints on the known physics coming from both earth-based and space experiments.  
In Table~\ref{tab:BPs} we present the masses and the couplings for the two benchmark points for the collider studies allowed by the Higgs data at the LHC \cite{Aaboud:2018xdt, Sirunyan:2018ouh} and the DM relic constraints \cite{Ade:2015xua}.


\begin{table}[t!]
	\centering\renewcommand*{\arraystretch}{1.3}\setlength{\tabcolsep}{18pt}
	\begin{tabular}{ |c|c|c|c|}
		\hline \hline
		\multirow{2}{*}{Parameters}& \multicolumn{2}{c|}{Benchmark points }\\
		\cline{2-3}
		&BP1&BP2\\\hline
$\lambda_{HT}$&0.67&-0.041\\\hline
$\kappa_{HT}$&55.85&299.70\\\hline
$\lambda_H$&0.26&0.31\\\hline
$\lambda_T$&0.57&0.67\\\hline
$v_T$&0.85&4.86\\\hline
$m_{h_D}$&125.17&125.09\\\hline
$m_{h_T}$&1411.75&1366.87\\\hline
$m_{a_P}$&1411.72&1365.79\\\hline
$m_{h_P^\pm}$&1411.89&1365.96\\\hline
$m_{h_T^\pm}$&1411.92&1367.02\\
		\hline
		\hline 
	\end{tabular}
	\caption{Benchmark points consistent with the Higgs data at the LHC and DM relic. The masses as well as the dimension-full parameter $v_T$ and $\kappa_{HT}$ are expressed in GeV.}  \label{tab:BPs}
\end{table}

\begin{table}[t!]
	\centering\renewcommand*{\arraystretch}{1.2}\setlength{\tabcolsep}{11pt}
	\begin{tabular}{ |c|c|c|c|c|c|c|c|c|}
		\hline
		\multicolumn{9}{|c|}{Branching Ratios}\\
		\hline
		\multirow{3}{*}{$h_T$}& \multicolumn{2}{c|}{$W^+W^-$}& \multicolumn{2}{c|}{$h_D \,h_D$}& \multicolumn{2}{c|}{$Z \,Z$}& \multicolumn{2}{c|}{$\bar t\, t$}\\
		\cline{2-9}
		&BP1&BP2&BP1&BP2&BP1&BP2&BP1&BP2\\\cline{2-9}
		&0.478&0.466&0.241&0.251&0.241&0.240&0.040&0.042\\\cline{1-9}
		
		\multirow{3}{*}{$h_T^+$}& \multicolumn{2}{c|}{$W^+Z$}& \multicolumn{2}{c|}{$W^+h_D$}& \multicolumn{2}{c|}{$\bar b\, t$}& \multicolumn{2}{c}{}\\
		\cline{2-7}
		&BP1&BP2&BP1&BP2&BP1&BP2&\multicolumn{2}{c}{}\\\cline{2-7}
		&0.479&0.472&0.479&0.483&0.042&0.044&\multicolumn{2}{c}{}\\\cline{1-7}
		\cline{1-7}
		\multirow{3}{*}{$h_P^+$}& \multicolumn{2}{c|}{$a_P\, (W^+)^*$}& \multicolumn{2}{c}{}& \multicolumn{2}{c}{}& \multicolumn{2}{c}{}\\
		\cline{2-3}
		&BP1&BP2&\multicolumn{6}{c}{}\\\cline{2-3}
		&1.000&1.000&\multicolumn{6}{c}{}\\\cline{1-3}
		\cline{1-3}
	\end{tabular}
	\caption{Branching ratios of $h_T$, $h_T^\pm$ and $h_P^\pm$ for the two benchmark points considered. The numerical values have been computed with \mgFull.}  \label{tab:BR}
\end{table}

We report in Table~\ref{tab:BR} the branching ratios of $h_T$, $h_T^\pm$ and $h_P^\pm$. The neutral triplet-like scalar $h_T$ decays dominantly in $W^+ W^-$, with a small difference between the two benchmark points. The next-to-leading decay channels are $h_T\to h_D h_D$ and $h_T\to Z Z$, with similar branching ratios. The decay into fermions of $h_T$ is less relevant, the highest branching ratio being $Br(h_T\to \bar t t)\sim0.04$. The charged triplet-like scalar has two competitive decay channels, $W^+ Z$ and $W^+ h_D$. The branching ratios for these two channels and for both the benchmark points is 
\begin{equation}
Br(h_T^+\to W^+Z)\sim Br(h_T^+\to W^+h_D)\sim0.48 .
\end{equation}
Finally the pure charged scalar $h_P^\pm$ has a single decay-channel, $h^\pm_P\to a_P(W^\pm)^*$, where the $W^\pm$ remains off-shell. Like the branching ratios, the total decay-widths of the triplet-like and pure charged scalars are also different. In fact, as we will see in Section~\ref{sec:longlived}, the pure charged scalar has a life-time large enough to be measured in experiments designed to detect long-lived particles. We've computed numerically the total decay-width of $h_T$, $h_T^\pm$ and $h_P^\pm$, for the two benchmark points considered, with \mgFull~\cite{Alwall:2014hca}. Their values are
\bea
&\Gamma_{h_T}^{BP1}=8.93\cdot10^{-2}\,{\rm GeV},\;\Gamma_{h_T}^{BP2}=2.69\,{\rm GeV},\\
&\Gamma_{h_T^\pm}^{BP1}=8.95\cdot10^{-2}\,{\rm GeV},\;\Gamma_{h_T^\pm}^{BP2}=2.70\,{\rm GeV},\\
&\Gamma_{h_P^\pm}^{BP1}=3.03\cdot10^{-16}\,{\rm GeV},\;\Gamma_{h_P^\pm}^{BP2}=3.32\cdot10^{-16}\,{\rm GeV}.
\eea

\begin{figure}[t!]
	\centering
	\mbox{\subfigure[]{\includegraphics[scale=.35]{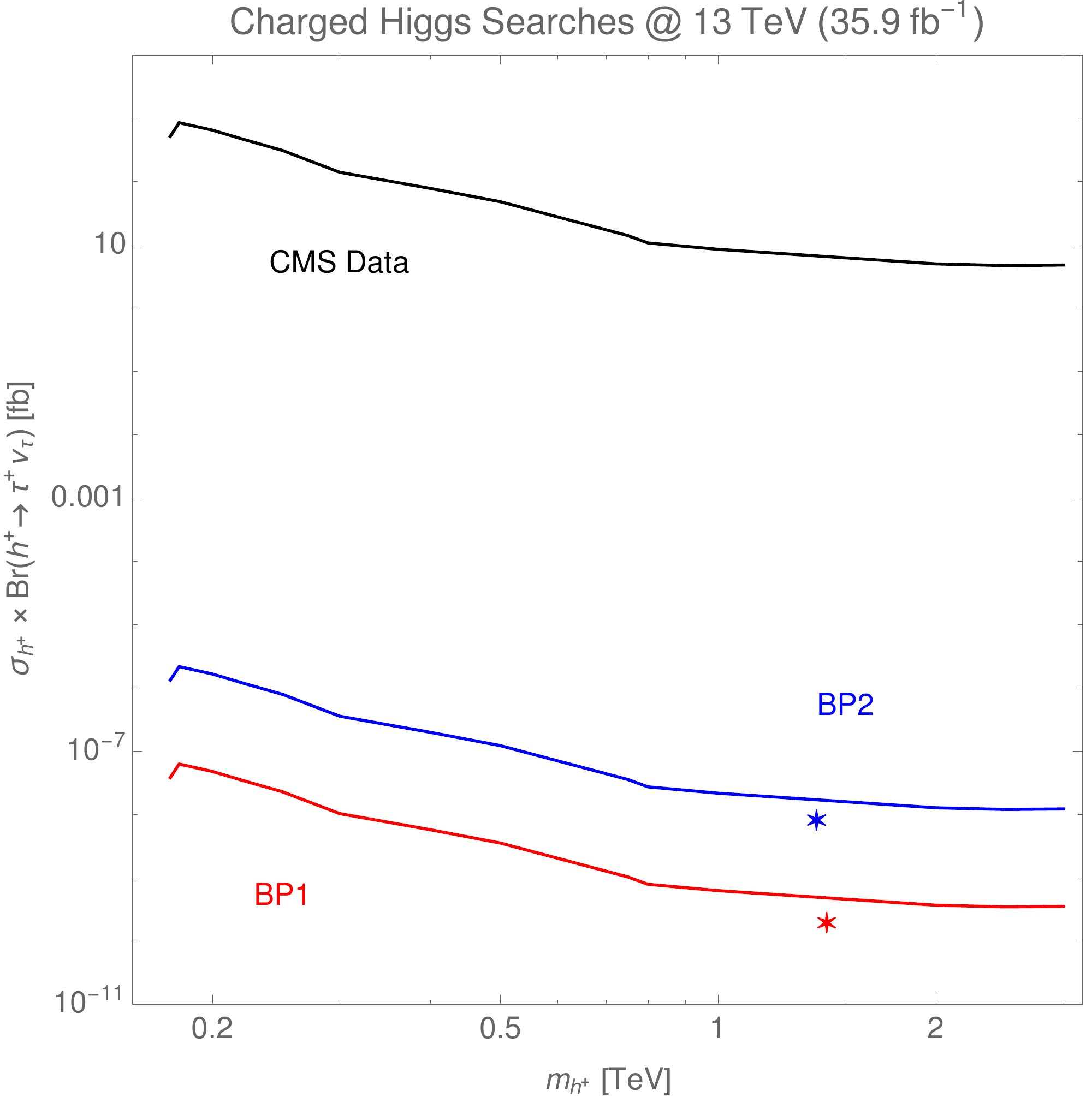}}\hspace{.5cm}\subfigure[]{\includegraphics[scale=.35]{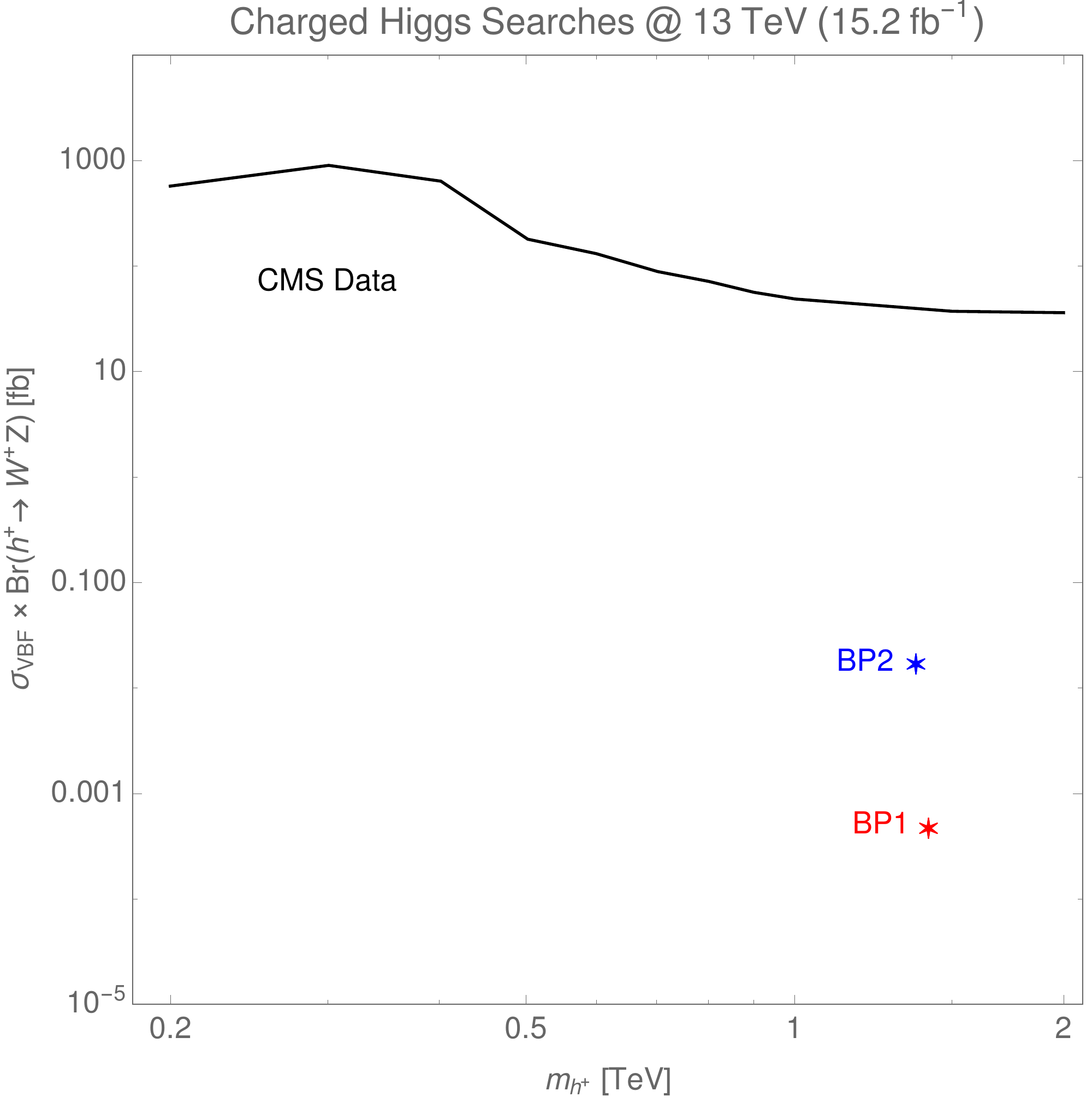}}
	}
	\caption{Heavy charged Higgs searches at the LHC. (a) Bound on $p\,p\to \bar t\, b\, h^+$ (with $h^+\to\tau^+\nu_\tau$) from CMS (black curve) \cite{Sirunyan:2019hkq}. Red/Blue line is the CMS bound times $(\mathcal R^C_{31})^2$ times the branching ratio $h_T^+\to\tau^+\nu_\tau$ for BP1/BP2. Red/Blue cross is $\sigma(\bar t\, b\, h_T^+)\times Br(h_T^+\to\tau^+\nu_\tau)$ computed with \mgFull~for BP1/BP2. (b) Bound on the vector-boson-fusion production of a charged Higgs decaying in $W\, Z$ bosons from CMS \cite{Sirunyan:2017sbn}. Red/Blue cross is $\sigma_{VBF}(h^+_T)\times Br(h_T^+\to W^+ Z)$ computed with \mgFull~for BP1/BP2.}\label{fig:cms_search}
\end{figure} 

It can be clearly seen the $h_T$ and $h_T^\pm$ will have prompt decays while $h_P^\pm$ will have displaced decays with a possibility of detection at the LHC \cite{CMS:2016isf,Sirunyan:2019nfw} and at MATHUSLA \cite{Lubatti:2019vkf}. However, the triplet-like charged Higgs $h_T^\pm$ with prompt decay will get constraints from the current LHC data. For this purpose we considered the Heavy charged Higgs searches at the LHC and present our results in Figure~\ref{fig:cms_search}. In Figure~\ref{fig:cms_search}(a) we show the bound on $p\,p\to \bar t\, b\, h^+$ (with $h^+\to\tau^+\nu_\tau$) from CMS (black curve) \cite{Sirunyan:2019hkq}. The Red and Blue lines represent the same bound computed with the branching ratios of $h_T^+\to\tau^+\nu_\tau$ of our benchmark points, where the cross-sections is suppressed by the doublet-triplet mixing $(\mathcal R^C_{31})^2$. Red and Blue crosses are $\sigma(\bar t\, b\, h_T^+)\times Br(h_T^+\to\tau^+\nu_\tau)$ computed with \mgFull~for BP1  and BP2 respectively.

Figure~\ref{fig:cms_search}(b) showcases the bound on the triplet-like production and decay modes, i.e. the vector-boson-fusion production of a charged Higgs decaying in $W\, Z$ bosons at the CMS \cite{Sirunyan:2017sbn}. Red and Blue crosses are $\sigma_{VBF}(h^+_T)\times Br(h_T^+\to W^+ Z)$ computed with \mgFull~for BP1 and BP2 respectively. Our benchmark points are allowed by both the constraints coming from Heavy charged Higgs searches at the LHC \cite{Sirunyan:2017sbn,Sirunyan:2019hkq}.

\subsection{Long-Lived Charged States}\label{sec:longlived}

In the cTSM the DM candidate is the pure pseudoscalar with $m_{DM}\approx1.35-1.60$ TeV. However, as we have shown in Eqs.~(\ref{mps}), (\ref{mcht}) and (\ref{mchp}), pseudoscalar, charged and neutral triplet state are almost degenerate in mass. The close degeneration in mass between $a_P$ and $h_P^\pm$ allow us to consider an interesting possibility. If fact even $h^\pm_P$ is a pure triplet state and this means that its couplings with the fermions are absent, similarly to the case of the pseudoscalar. However, the coupling $a_P\,h^\pm_P\,W^\mp$ is non-zero and hence the decay $h^\pm_P\to a_P\,W^\pm$ is possible. In fact the charged component of a multiplet can be slightly heavier than its neutral counterpart even at the tree-level and gets additional contribution to the mass splitting of $\mathcal{O}(10^2)$ MeV at one-loop \cite{Cirelli:2009uv}. The total mass splitting is still quite small and hence only the three-body decay of the charged Higgs bson $h^\pm_P$ is possible. This will allow us to consider $h_P^\pm$ as a Long-Lived (LL) heavy state and its lifetime is in the range testable by the proposed experiment MATHUSLA \cite{Lubatti:2019vkf}.

The partial decay width of $h^\pm_P$ in $a_P\,W^\pm$ is given by \cite{Djouadi:1995gv}
\bea
\frac{d\Gamma}{dx_1dx_2}(h^\pm_P\to a_P W^{*\pm}\to a_P f f^\prime)=\frac{9}{8\pi^3}G_F^2\, m_W^4\, m_{h_P^\pm}\,F_{a_P W^\pm}(x_1, x_2),
\eea
where
\bea
F_{XY}(x_1,x_2)=\frac{(1-x_1)(1-x_2)-\kappa_X}{(1-x_1-x_2-\kappa_X+\kappa_Y)^2+\kappa_Y\gamma_Y}
\eea
and, for the decay $A\to XY^*$, $\kappa_{X,Y}=m^2_{X,Y}/m^2_A$, $\gamma_Y=\Gamma^2_Y/m^2_A$. $G_F$ is the Fermi constant of the weak interaction. 
For the quark finalstates can give rise to charged pion finalstate with the decay width given by
\bea
\Gamma_\pi= \frac{2}{\pi}G_F^2\, V_{ud}^2\, \Delta m^3\, f_\pi\, \sqrt{1- \frac{m_\pi^2}{\Delta m^2}},
\eea
where $f_\pi= 131$ MeV, $\mathbf{V}$ is the CKM matrix and $\Delta m\sim166$ MeV \cite{Cirelli:2009uv, Chun:2009mh}.
 Considering that our DM candidate has a mass $m_{DM}\sim1.5$ TeV and that the mass splitting between the charged and pseudoscalar Higgs boson is $\Delta m\sim166$ MeV, we have that the lifetime of the charged Higgs boson is
\bea
\tau_{h_P^\pm}=\mathcal{O}(10^{15})\, \textrm{GeV}^{-1}=\mathcal{O}(1)\,\textrm{m} .
\eea
This is in agreement with the numerical computation of the total decay width of $h_P^\pm$ with \mgFull. However, for the general case of the potential, i.e. with Eq.~\ref{extrapot}, the mass difference between the pure charged Higgs and pure pseudoscalar is $(2\,\lambda_{HT}^{(2)}+\lambda_{T'}/2)\,v_T^2$ (see Eq.\ref{massdiff}). This mass splitting can be of $\mathcal{O}(1)$ GeV, giving rise to displacement of a few mm only.

\begin{figure}[t!]
	\centering
	\mbox{\subfigure[]{\includegraphics[scale=.35]{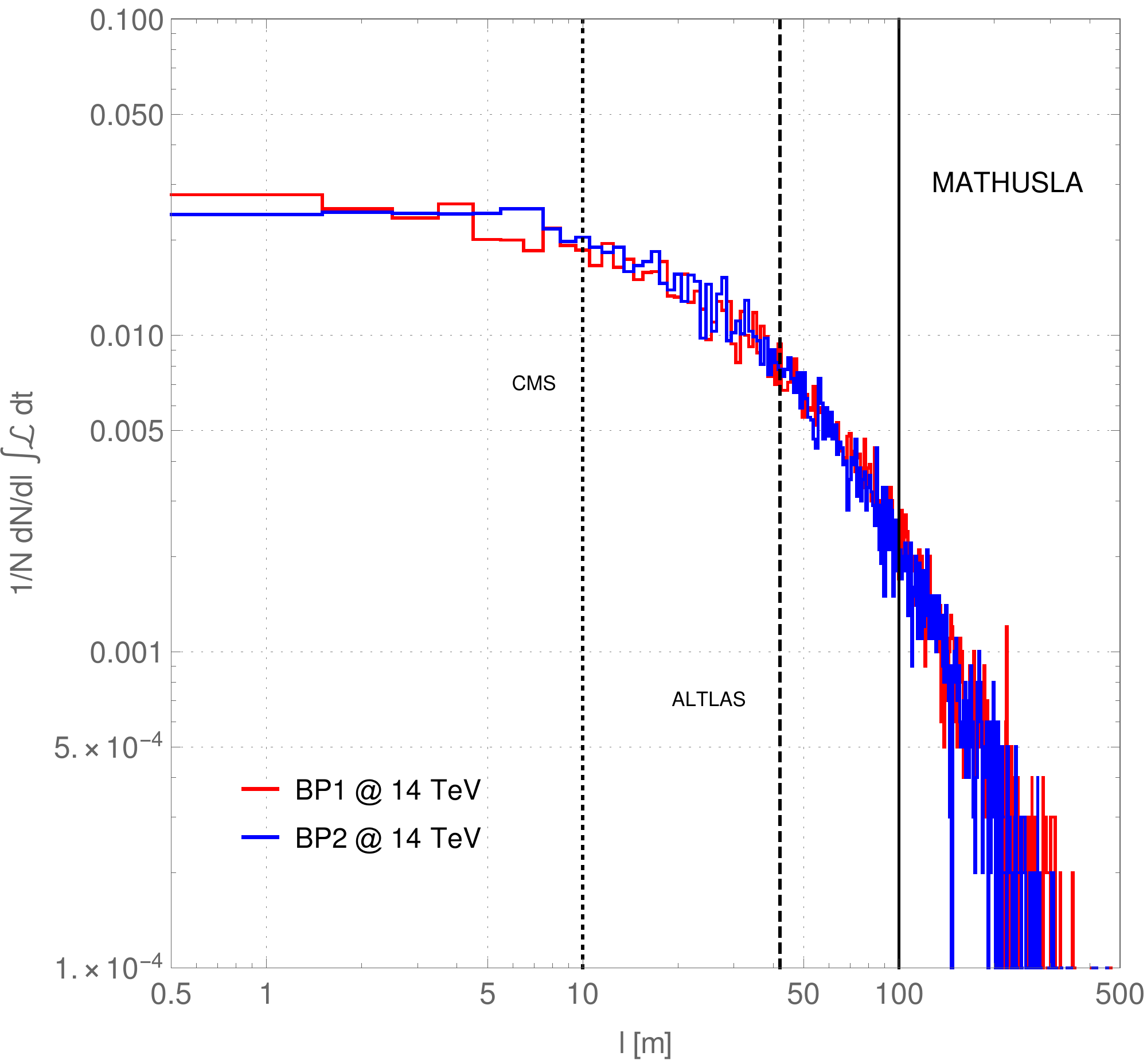}}\hspace{.5cm}\subfigure[]{\includegraphics[scale=.35]{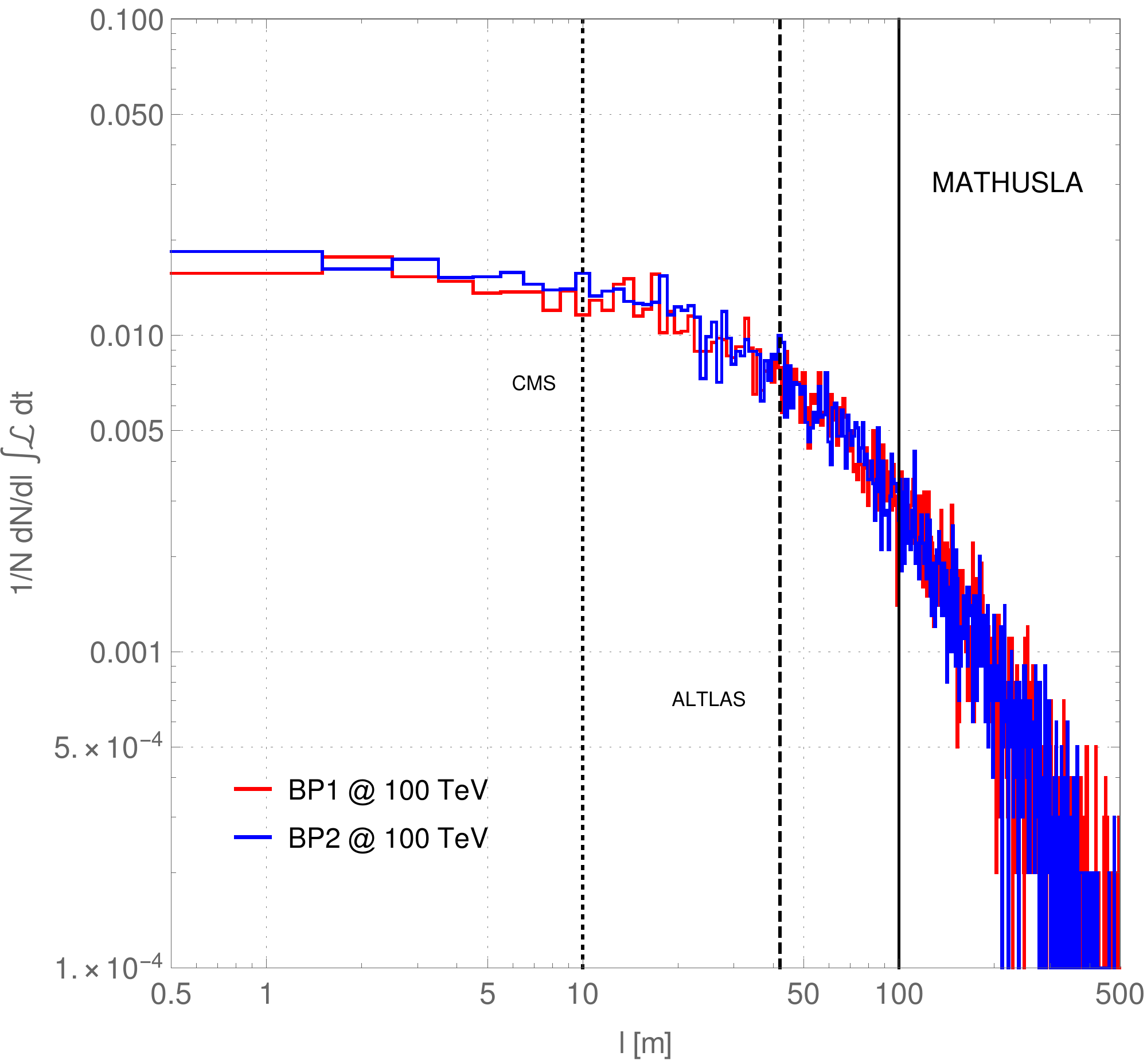}}
	}
	\caption{Decay length distribution for the LL particle $h_P^\pm$. We have considered the pair-production $pp\to h_P^+ h_P^-$ at hadron colliders, with (a) $\sqrt s = 14$ TeV and (b) $\sqrt s = 100$ TeV. We mark with a dashed(dotted) line the ATLAS(CMS) upper limit for LL searches. The solid black line is the MATHUSLA lower limit \cite{Lubatti:2019vkf}.}\label{fig:longlived}
\end{figure} 

In Figure~\ref{fig:longlived} we present the displaced decay length versus normalized  number of events of the pure charged Higgs boson.  Here we have considered the pair-production at the LHC with centre-of-mass energies of 14 TeV in Figure~\ref{fig:longlived}(a) and 100 TeV in Figure~\ref{fig:longlived}(b) respectively. The vertical lines isolate different regions of detectability, i.e. CMS $\sim 10$ m, ATLAS $\sim 40$ m and MATHUSLA $\sim 100-500$ m respectively. The decay length has been computed with \texttt{Pythia8.2} \cite{Sjostrand:2014zea}. The observations of the displaced charged Higgs boson is very similar to the ones we saw in case of the real triplet \cite{Jangid:2020qgo} with a difference that now we have an additional triplet-like charged Higgs boson $h_T^\pm$ which gives prompt decays in the similar mass range. 

In the context of supersymmetry such triplet-like charged Higgs bosons mix with the doublet ones. This is also true for the massive pseudoscalar boson that lose its purity in terms of gauge eigenstates \cite{Bandyopadhyay:2014vma, Bandyopadhyay:2015ifm}. In this case the pseudoscalar cannot become the dark matter candidate.  The charged Higgs bosons in these cases can give rise to the triplet-like signature decaying to $ZW^\pm$, which is proportional to the square of the triplet VEV. Such decays are however prompt ones. Interesting scenarios appear when one considers the charged Higgs boson super-partner, i.e. the chargino, which can give rise to displaced decays \cite{SabanciKeceli:2018fsd}. The displaced Higgs boson decays can also appear in various SUSY scenarios \cite{Bandyopadhyay:2010wp,Bandyopadhyay:2010cu}.

\subsection{Self-Couplings of the Higgs boson(s)}

The self-couplings of the neutral scalar Higgs boson are important ingredients for a clear understanding of the EWSB mechanism. In the SM there is only one quartic self coupling,  $\lambda_{\textrm{SM}}$, which encodes all the information about the scalar potential. In models with an enlarged scalar sector the situation can be very different. We have no more just one dimensionless parameter in the potential, i.e. $\lambda_{\textrm{SM}}$, and the relation $\lambda^{(4)}_{\textrm{SM}}= \lambda^{(3)}_{\textrm{SM}}/v$ does not hold in general.

\begin{figure}[t]
	\centering
	\includegraphics[scale=.65]{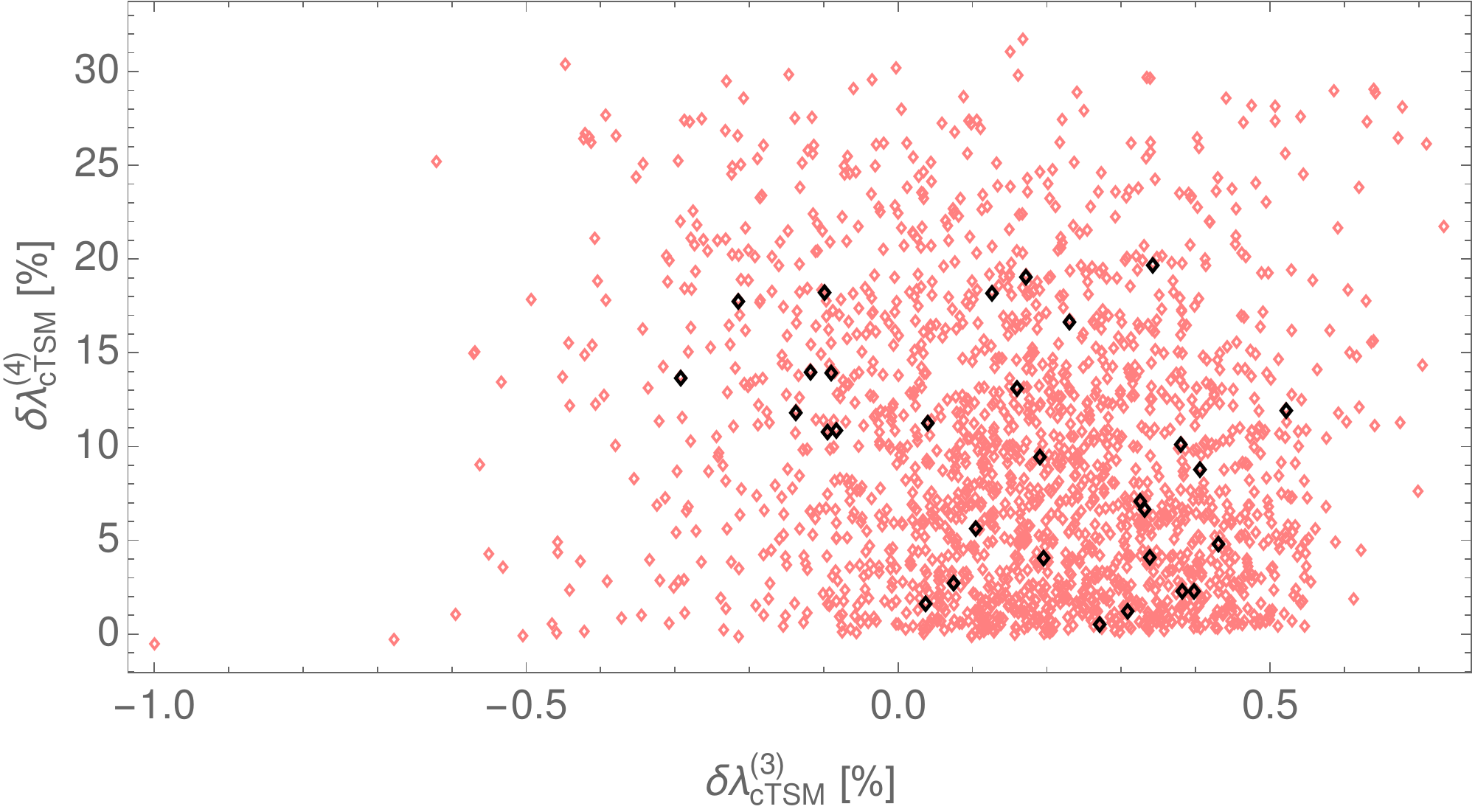}
	\caption{Correlation plot of the trilinear and quartic couplings $\lambda^{(3)}_{cTSM}$ and $\lambda^{(4)}_{cTSM}$ in the cTSM. The SM values are marked with the star. The black points correspond to the pseudoscalar satisfying the relic density for the DM. }\label{triquarF}
\end{figure}

In the cTSM the trilinear and quartic couplings of the SM-like Higgs boson ($h_D$) are expressed by
\bea
\lambda^{(3)}_{cTSM}&\equiv& g_{h_D h_D h_D}=3 \lambda_H\, v\, (\mathcal{R}_{11}^S)^3-3 (\kappa_{HT}-\lambda_{HT}\,v_T/2)(\mathcal{R}_{11}^S)^2\mathcal{R}_{12}^S\label{trilH}\\
&&\qquad\qquad\qquad+3 \lambda_{HT}\, v/2 \,\mathcal{R}_{11}^S(\mathcal{R}_{12}^S)^2+3(2\lambda_{T}+\lambda_{T'}) v_T/2(\mathcal{R}_{12}^S)^3, \nn\\
\lambda_{cTSM}^{(4)}&\equiv& g_{h_D h_D h_D h_D}=3 \lambda_H (\mathcal{R}_{11}^S)^4+3 \lambda_{HT}(\mathcal{R}_{11}^S)^2(\mathcal{R}_{12}^S)^2+3/2 (2\lambda_{T}+\lambda_{T'}) (\mathcal{R}_{12}^S)^4.\label{quartH}
\eea
We can see from Eqs.~(\ref{trilH}) and (\ref{quartH}) that $\lambda_{cTSM}^{(3)}=\lambda_{SM}^{(3)}(\mathcal{R}_{11}^S)^3+\cdots$ and similarly $\lambda_{cTSM}^{(4)}=\lambda_{SM}^{(4)}(\mathcal{R}_{11}^S)^4+\cdots$. We see that even in the SM-like quartic and cubic couplings will have contamination from the triplet parameters. To illustrate that we plot in Figure~\ref{triquarF} the correlation plot between $\delta\lambda^{(3)}_{cTSM}$ and $\delta\lambda^{(4)}_{cTSM}$, defined as
\begin{equation}
\delta\lambda_{cTSM}^{(i)}\equiv \frac{\lambda_{cTSM}^{(i)} - \lambda_{SM}^{(i)}}{\lambda_{SM}^{(i)}}.
\end{equation}
We have marked in black the points for which the mass of the DM candidate $a_P$ is compatible with the constraint on the relic density obtained from Planck, as previously discussed. For these points the difference in the trilinear self-coupling of the SM-like Higgs boson $h_D$ is below 1\%. Such a small deviation from the SM prediction for the trilinear coupling of the Higgs boson cannot be observed at the proposed future hadron colliders. In fact it is expected that the Future Circular Collider (FCC) \cite{Benedikt:2018csr,Abada:2019lih,Abada:2019zxq} will provide a measurement of the trilinear coupling with $\mathcal O(5\%)$ accuracy \cite{Abada:2019lih,DiMicco:2019ngk}. The quartic coupling exhibit a maximum deviation of 20\% for the points that are satisfying the constraint on the relic density. In general the expected constraint at future colliders for the quartic coupling is looser than the trilinear one \cite{Papaefstathiou:2015paa,Contino:2016spe,Fuks:2017zkg,Papaefstathiou:2019ofh}. A deviation of $\sim20\%$ will not be visible at proposed next-generation of hadron or lepton colliders \cite{Bizon:2018syu,Borowka:2018pxx,Chiesa:2020awd}.

\section{cTSM at Colliders}\label{sec:collider}

We have already pointed out that the couplings of the triplet states with the fermions are suppressed if not absent, as in the case of $a_P, h^\pm_P$. At the hadron colliders, like the LHC, this will make their search quite challenging. The single production of triplet-like states at hadron colliders ($h_T$ and $h_T^\pm$) will proceed via quark-fusion but with a suppression of order $(v_T/v)^2$ with respect to their SM counterpart. Similar considerations will hold for the triplet-like states pair-production.

The situations can be different at very-high-energy lepton collider. Let us consider as an example a multi-TeV muon collider. If the centre-of-mass energy is sufficiently high $(\sqrt s \gsim 10\,\rm{TeV})$ the muon collider became effectively a vector-boson collider \cite{Costantini:2020stv}. A muon collider faces many problems concerning its functioning. The ultimate reason for these issues is the fact that muon decay and their lifetime is also short. Nonetheless the High-Energy-Physics community has put a lot of interest in this option for the future colliders \cite{Palmer:1996gs,Ankenbrandt:1999cta,Palmer:2014nza,Antonelli:2015nla,Delahaye:2019omf,Bartosik:2019dzq,Strategy:2019vxc,EuropeanStrategyGroup:2020pow,Chiesa:2020awd,Costantini:2020stv,Capdevilla:2020qel,Han:2020uid,Long:2020wfp,Han:2020pif,Han:2020uak}.     
This is justified by the astonishing possibilities that a very-high-energy muon collider offer from the physics side. It will be a facility for a test of the SM at high-precision level but also a discovery machine for BSM physics \cite{Costantini:2020stv}. 

We now discuss the cTSM at colliders. For this purpose we use \mgFull~\cite{Alwall:2014hca} and computed the cross-section for the most relevant production channels of the extra scalars $a_P$, $h_T$, $h_P^\pm$ and $h_T^\pm$.

\subsection{Hadron Colliders}

\begin{figure}[t!]
	\centering
	\includegraphics[scale=1.3]{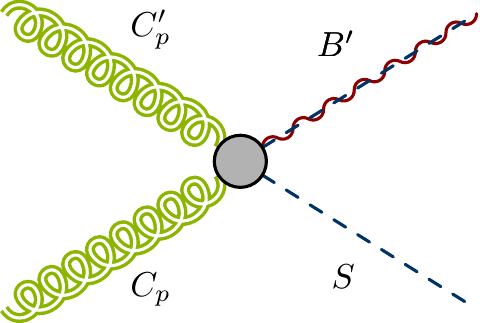}
	\caption{Schematic production process of a scalar ($S$) and a boson ($B'$) at an hadron collider. We depict with a green spring a colored parton ($C_p$). This can be either a quark or a gluon. The scalar $S$ (blue-dashed line) can be either a neutral or a charged one and the boson $B'$ (dashed-blue/wavy-red line) can be either a scalar or a vector boson.}\label{fig:pp}
\end{figure}
\begin{table}[t!]
	\centering\renewcommand*{\arraystretch}{1.2}\setlength{\tabcolsep}{10pt}
	\begin{tabular}{ |l|c|c||c|c|}
		\hline \hline
		\multirow{3}{*}{Production modes}& \multicolumn{4}{c|}{$\sigma$ [fb] }\\
		\cline{2-5}
		& \multicolumn{2}{c||}{$\sqrt s = 14$ TeV}& \multicolumn{2}{c|}{$\sqrt s=100$ TeV}\\
		\cline{2-5}
		&BP1&BP2&BP1&BP2\\\hline
		$\quad p\;p\to h_T $&$6.7\cdot10^{-7}$&$2.7\cdot10^{-5}$&$8.4\cdot10^{-5}$&$3.2\cdot10^{-3}$\\\hline
		$\quad p\;p\to h_T^\pm$&$8.2\cdot10^{-7}$&$3.2\cdot10^{-5}$&$9.5\cdot10^{-5}$&$3.5\cdot10^{-3}$\\\hline
		\hline
		$\quad p\;p\to h_T\; h_T$&$2.3\cdot10^{-7}$&$1.6\cdot10^{-8}$&$4.3\cdot10^{-4}$&$2.7\cdot10^{-5}$\\\hline
		$\quad p\;p\to a_P\;a_P$&$2.2\cdot10^{-7}$&$1.1\cdot10^{-9}$&$4.2\cdot10^{-4}$&$1.8\cdot10^{-6}$\\\hline
		$\quad p\;p\to h^+_T\; h^-_T$&$3.9\cdot10^{-3}$&$4.9\cdot10^{-3}$&$1.3\cdot10^0\cdot$&$1.4\cdot10^0$\\\hline
		$\quad p\;p\to h^+_P\;h^-_P$&$3.9\cdot10^{-3}$&$4.9\cdot10^{-3}$&$1.3\cdot10^0\cdot$&$1.4\cdot10^0$\\\hline
		\hline 
		$\quad p\;p\to h_D\;h_T$&$1.5\cdot10^{-5}$&$5.4\cdot10^{-4}$&$5.1\cdot10^{-3}$&$1.8\cdot10^{-1}$\\\hline
		$ \quad p\;p\to h_D\;h_T^\pm$&$1.7\cdot10^{-6}$&$6.7\cdot10^{-5}$&$1.1\cdot10^{-4}$&$4.1\cdot10^{-3}$\\\hline
		\hline
		$\quad p\;p\to h_T\; Z $&$1.3\cdot10^{-6}$&$5.0\cdot10^{-5}$&$1.0\cdot10^{-4}$&$3.7\cdot10^{-3}$\\\hline
		$\quad p\;p\to h_T\;W^\pm$&$1.9\cdot10^{-6}$&$7.3\cdot10^{-5}$&$1.2\cdot10^{-4}$&$4.3\cdot10^{-3}$\\\hline
		$\quad p\;p\to h^\pm_T\; Z$&$1.9\cdot10^{-6}$&$7.5\cdot10^{-5}$&$1.2\cdot10^{-4}$&$4.4\cdot10^{-3}$\\\hline
		$\quad p\;p\to h^+_T\;W^-$&$2.4\cdot10^{-5}$&$9.1\cdot10^{-4}$&$4.2\cdot10^{-2}$&$1.5\cdot10^0$\\\hline
		\hline
		$\quad p\;p\to h_T\;p\;p'$&$3.1\cdot10^{-7}$&$1.4\cdot10^{-5}$&$7.9\cdot10^{-5}$&$3.9\cdot10^{-3}$\\\hline
		$\quad p\;p\to h_T^\pm\;p\;p'$&$3.6\cdot10^{-7}$&$1.4\cdot10^{-5}$&$8.5\cdot10^{-5}$&$3.1\cdot10^{-3}$\\\hline
		\hline
	\end{tabular}
	\caption{Various BSM production processes from $pp$ collisions and VBF at hadron colliders. The c.o.m. energy considered are $\sqrt s=14$ TeV as benchmark energy at the LHC and $\sqrt s = 100$ TeV as benchmark energy at FCC.}  \label{tab:hadronXS}
\end{table}

First we consider the production cross-sections at hadron collider, i.e. at the LHC. In Figure~\ref{fig:pp} we represent (very) schematically a production process at an hadron collider. In order to be as generic as possible, we name a colored parton $C_p$ (this can be either a quark or a gluon), marked by a green spring.  Moreover $S$ is a scalar boson (either charged or neutral), marked in blue, and $B'$ is either a scalar or a vector boson (we chose a mixed line to depict them). Assuming that $B'$ may or may not be produced, Figure~\ref{fig:pp} represents the various production processes listed in Table~\ref{tab:hadronXS}, apart the last two. These are the single production of $h_T/h_T^\pm$; the pair production of $a_P$, $h_T$, $h_P^\pm$ and $h_T^\pm$; the associated production of $h_T/h_T^\pm$ with the SM-like Higgs boson or the massive vector bosons.

The production cross-sections listed in Table~\ref{tab:hadronXS} have been computed for $\sqrt s =14, 100$ TeV with \mgFull. We've used the \texttt{nnpdf2.3qed} parton distribution functions \cite{Ball:2013hta} with  $\mu_F=m_Z$. At $\sqrt s = 14$ TeV, considered as the benchmark energy for the LHC, the production cross-sections for the BSM states cover the range $10^{-7} - 10^{-3}$ fb for BP1 and $10^{-9}-10^{-3}$ fb for BP2. These cross-sections are too low to have any chance of discovery at the LHC. The benchmark energy considered for the Future Circular Collider (FCC) is $\sqrt s = 100$ TeV \cite{Mangano:2017tke}. By inspection of Table~\ref{tab:hadronXS} we can see that in this scenario the production cross-sections span over the range $10^{-5}-10^0$ for BP1 and $10^{-6}-10^0$ for BP2. However, although the production cross-section is enhanced by $2/3$ orders of magnitude from the $\sqrt s=14$ TeV to the $\sqrt s=100$ TeV case but still too feeble to be resolved from SM backgrounds in general. Thus one has to look for multi-lepton final-states to win over the SM backgrounds \cite{Bandyopadhyay:2014vma}.

\subsection{Muon Collider}

\begin{figure}[t!]
	\centering
	\includegraphics[scale=1.0]{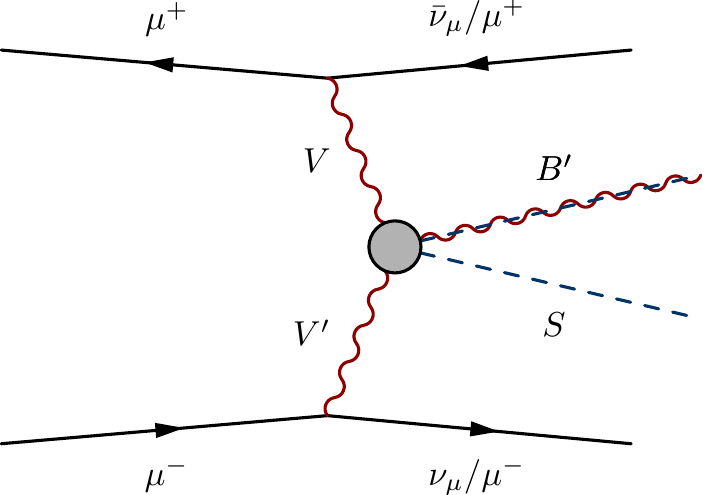}
	\caption{Schematic production process via VBF of a scalar ($S$) and a boson ($B'$) at a muon collider. We depict with a wavy-red line a gauge boson ($\gamma,\,Z$ or $W^\pm$). The scalar $S$ (blue-dashed line) can be either a neutral or a charged one and the boson $B'$ (dashed-blue/wavy-red line) can be either a scalar or a vector boson.}\label{fig:muoncoll}
\end{figure} 
\begin{table}[t!]
	\centering\renewcommand*{\arraystretch}{1.2}\setlength{\tabcolsep}{10pt}
	\begin{tabular}{ |l|c|c||c|c|}
		\hline \hline
		\multirow{3}{*}{Production modes}& \multicolumn{4}{c|}{$\sigma$ [fb] }\\
		\cline{2-5}
		& \multicolumn{2}{c||}{$\sqrt s = 14$ TeV}& \multicolumn{2}{c|}{$\sqrt s=30$ TeV}\\
		\cline{2-5}
		&BP1&BP2&BP1&BP2\\\hline
		$\mu^+\mu^-\,\to h_T \;\nu_\mu\bar\nu_\mu $&$1.8\cdot10^{-2}$&$6.2\cdot10^{-1}$&$2.9\cdot10^{-2}$&$9.6\cdot10^{-1}$\\\hline
		$\mu^+\mu^-\to h_T^+\;\mu^-\bar\nu_\mu$&$5.3\cdot10^{-3}$&$1.8\cdot10^{-1}$&$8.4\cdot10^{-3}$&$2.8\cdot10^{-1}$\\\hline
		\hline
		$\mu^+\mu^-\to h_T\; h_T\;\nu_\mu\bar\nu_\mu$&$1.9\cdot10^{-2}$&$2.0\cdot10^{-2}$&$4.8\cdot10^{-2}$&$5.1\cdot10^{-2}$\\\hline
		$\mu^+\mu^-\to a_P\;a_P\;\nu_\mu \bar\nu_\mu$&$1.8\cdot10^{-2}$&$2.0\cdot10^{-2}$&$4.7\cdot10^{-2}$&$5.0\cdot10^{-2}$\\\hline
		$\mu^+\mu^-\to h^+_T\; h^-_T\;\nu_\mu \bar\nu_\mu$&$1.3\cdot10^{-2}$&$1.4\cdot10^{-2}$&$3.4\cdot10^{-2}$&$3.6\cdot10^{-2}$\\\hline
		$\mu^+\mu^-\to h^+_P\;h^-_P\;\nu_\mu \bar\nu_\mu$&$1.3\cdot10^{-2}$&$1.4\cdot10^{-2}$&$3.4\cdot10^{-2}$&$3.6\cdot10^{-2}$\\\hline
		\hline 
		$\mu^+\mu^-\to h_D\;h_T\;\nu_\mu \bar\nu_\mu$&$1.6\cdot10^{-4}$&$5.7\cdot10^{-3}$&$3.7\cdot10^{-4}$&$1.3\cdot10^{-2}$\\\hline
		$ \mu^+\mu^-\to h_D\;h_T^+\;\mu^-\bar\nu_\mu$&$4.8\cdot10^{-5}$&$1.6\cdot10^{-3}$&$1.1\cdot10^{-4}$&$3.8\cdot10^{-3}$\\\hline
		\hline
		$\mu^+\mu^-\to h_T\; Z\;\nu_\mu \bar\nu_\mu$&$7.7\cdot10^{-4}$&$2.6\cdot10^{-2}$&$1.7\cdot10^{-3}$&$5.6\cdot10^{-2}$\\\hline
		$\mu^+\mu^-\to h_T\;W^+\mu^-\bar\nu_\mu$&$4.1\cdot10^{-4}$&$1.4\cdot10^{-2}$&$1.0\cdot10^{-3}$&$3.4\cdot10^{-2}$\\\hline
		$\mu^+\mu^-\to h^+_T\; Z\;\mu^-\bar\nu_\mu$&$1.4\cdot10^{-4}$&$4.8\cdot10^{-3}$&$3.6\cdot10^{-4}$&$1.2\cdot10^{-2}$\\\hline
		$\mu^+\mu^-\to h^+_T\;W^-\;\nu_\mu \bar\nu_\mu$&9.7$\cdot10^{-4}$&$3.2\cdot10^{-2}$&$1.9\cdot10^{-3}$&$6.1\cdot10^{-2}$\\\hline
		\hline
	\end{tabular}
	\caption{Various BSM production processes via $W$-boson-fusion or $W^+\,Z/\gamma^*$-boson-fusion at a multi-TeV muon collider.}  \label{tab:muonXS}
\end{table}

Let us now consider another possibility for the future colliders. Lepton machines are usually thought to be precision machine, suitable for testing the know features of the SM. Although this is certainly true for low-energy $e^+e^-$ colliders, the possibility to search for BSM physics at a lepton collider has been considered and partially exploited in the recent years. In particular, the possibility of a circular $\mu^+ \mu^-$ collider running at several-to-many TeV has attracted the community. 
Despite the technical issues, ultimately related to the short lifetime of the muons, a circular $\mu^+\mu^-$ collider has many dream-like features. Among the other, we mention explicitly the huge advantage in terms of parton luminosity compared to a $pp$ collider running at the same energy \cite{Costantini:2020stv}.

In Figure~\ref{fig:muoncoll} we depict, very schematically, a production process via vector-boson-fusion at a muon collider. Here $V,\,V'$ are vector bosons, either $W$, $Z$ or $\gamma$, whereas $S$ and $B'$ are the same of Figure~\ref{fig:pp}. With these definitions Figure~\ref{fig:muoncoll} represents the various production processes listed in Table~\ref{tab:muonXS} computed by \mgFull~at the tree-level with centre of mass energies of 14 and 30 TeV. 

By comparing Table~\ref{tab:muonXS} and Table~\ref{tab:hadronXS} we can see that, concerning the single production,  the cross-section at a multi-TeV collider highly overcome the ones at an hadron collider ate the same energy. If we take $\sqrt s_p=\sqrt s_\mu=14$ TeV
\begin{equation}
\frac{\sigma^{14\,\rm TeV}_\mu(X)}{\sigma^{14\,\rm TeV}_p(X)}=10^4-10^2 .
\end{equation}
Moreover we see that $\sigma^{14\,\rm TeV}_\mu(X)\gsim10^2\,\sigma^{100\,\rm TeV}_p(X)$. Similar arguments hold for the neutral pair-production. A hadron collider at 100 TeV is competitive with a 14 TeV muon collider for the charged scalars pair-production and the associated production of $h_T/h_T^\pm$ with SM(-like) particles.

\begin{figure}[t!]
	\centering
	\mbox{
		\subfigure[]{\includegraphics[scale=.37]{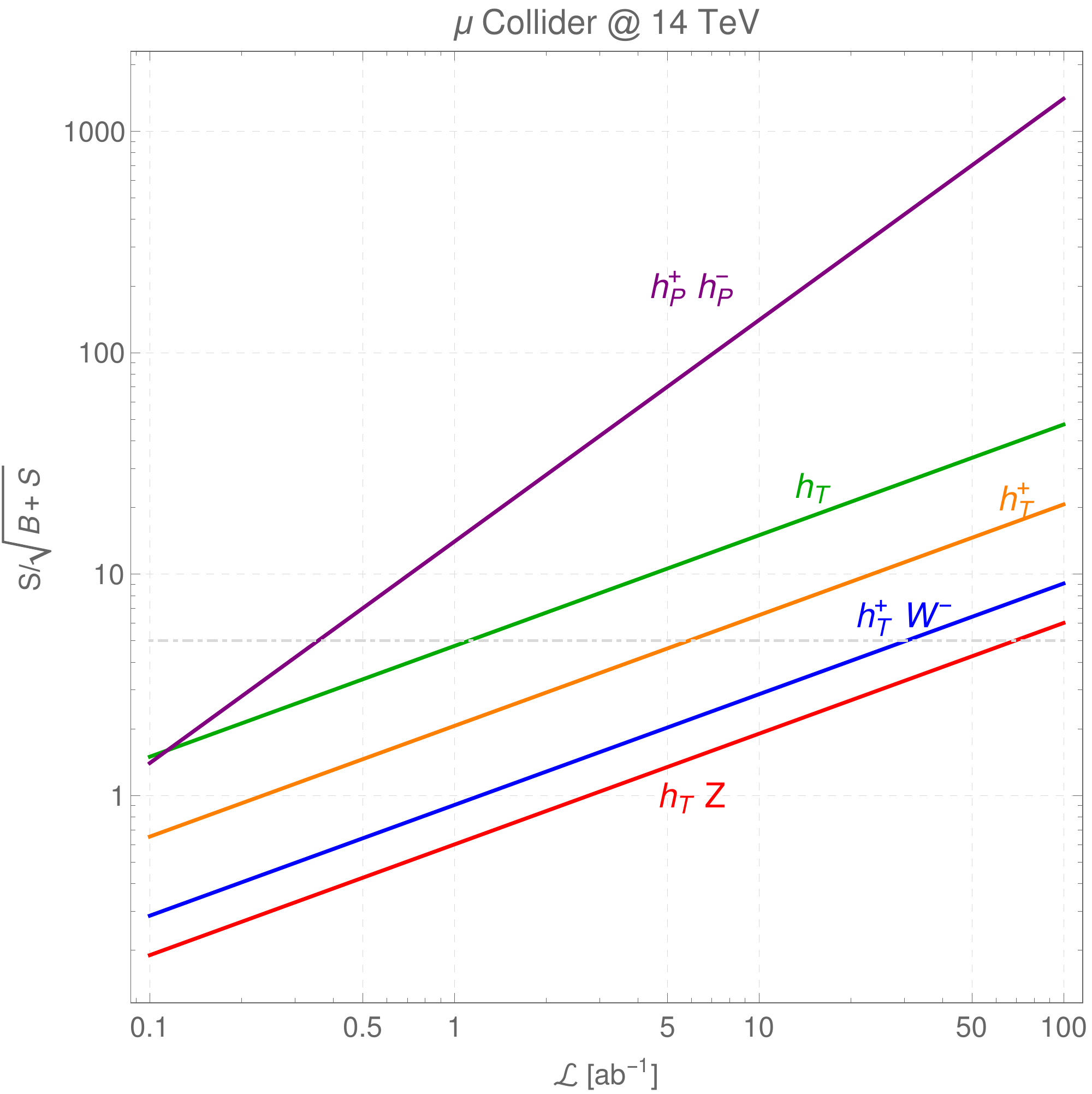}}
		\subfigure[]{\includegraphics[scale=.355]{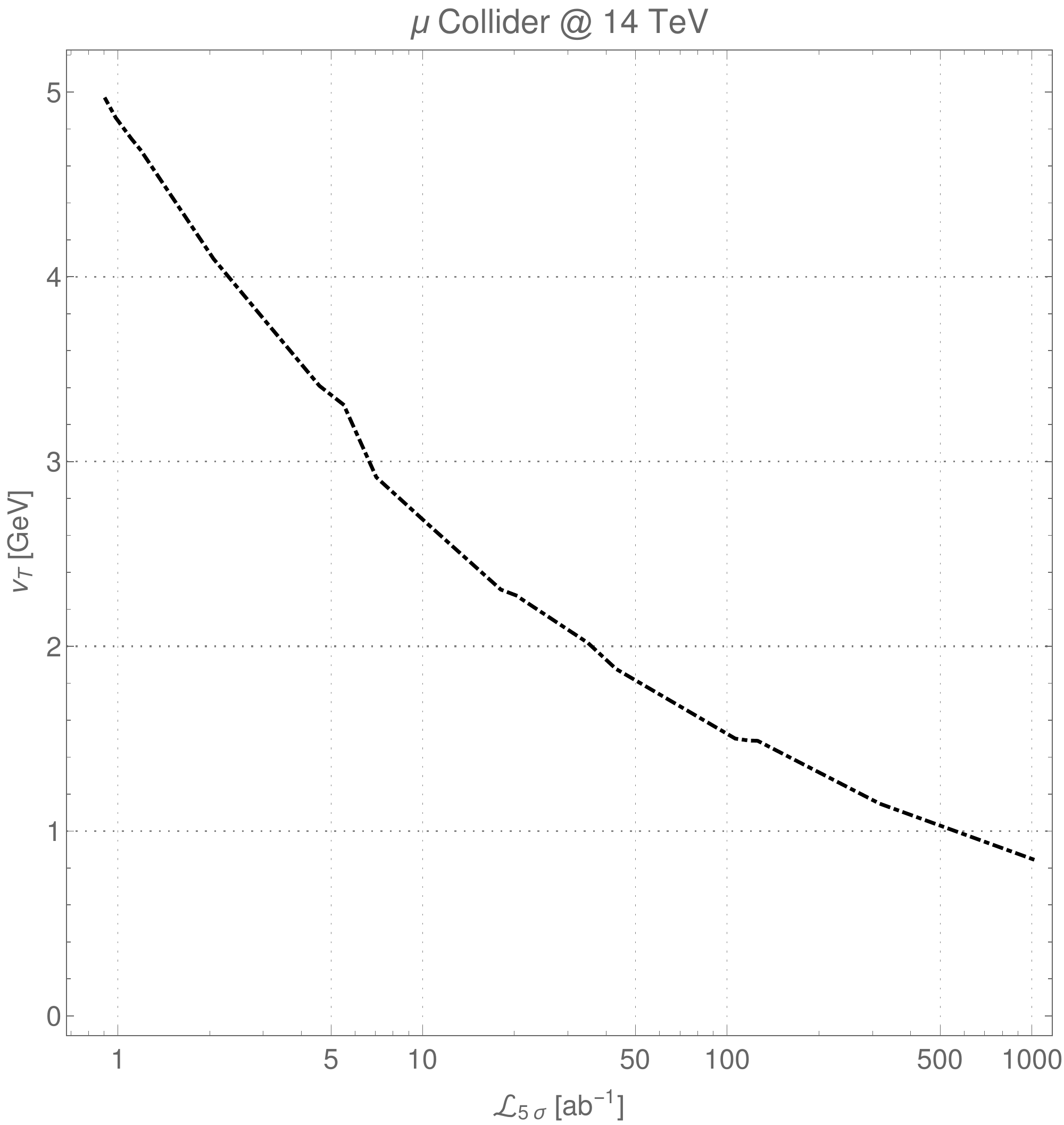}}
	}
	\caption{(a) Significance vs luminosity for the $h_T$, $h_T^\pm$, $h_T^+\, W^-$, $h_T\, Z$ and $h_P^+ h_P^-$ production processes through VBF at a 14 TeV muon collider. The production cross-sections are multiplied by the branching ratios $Br(h_T\to W^+W^-)$ or $Br(h_T^+\to W^+Z)$, depending on the channel considered. The background considered is the VBF production of $W^+ W^- Z$ in the SM, with $M(W^+ W^-)=m_{h_T}$ or $M(W^+ Z)=m_{h_T^+}$ for BP2. The process $h_P^+ h_P^-$ is considered background-free because $h_P^\pm$ is a LL state.  (b) Plot of $v_T$ vs discover luminosity obtained from the VBF production of $h_T$ at a 14 TeV muon collider. The black dot-dashed line is obtained from the points that satisfy the Planck constraint on the DM relic density.}\label{fig:zhtsig}
\end{figure} 

We have also considered the physics reach for some of the listed processes at a muon collider. These results are presented in Figure~\ref{fig:zhtsig}. In Figure~\ref{fig:zhtsig}(a) we plot the significance as function of the luminosity for the $h_T$, $h_T^\pm$, $h_T^+\, W^-$, $h_T\, Z$ and $h_P^+h_P^-$ production processes through VBF at a 14 TeV muon collider. In the definition of the significance, $\sigma=S/\sqrt{S+B}$, $S$ and $B$ stand for the number of events for the signal and the background respectively. For each line of Figure~\ref{fig:zhtsig}(a) we have considered the production cross-section relative to BP2 and multiplied by a branching ratio. To be specific, for the green line signal and background are given by
\begin{align}
S:& \;\sigma(\mu^+\mu^-\,\to h_T \;\nu_\mu\bar\nu_\mu) \times Br(h_T\to W^+W^-) \cdot \mathcal{L},\\
B:& \;\sigma(\mu^+\mu^-\,\to W^+W^- \;\nu_\mu\bar\nu_\mu) \cdot \mathcal{L},
\end{align}
with $M(W^+ W^-)=m_{h_T}\pm5$ GeV.

A similar strategy is applied to the single production of $h_T^\pm$ and the pair-production $h_T Z$ and $h_T^+W^-$. For the charged scalar Higgs $h_T^\pm$ we have considered the branching ratio $Br(h_T^+\to W^+Z)$. This give us a conservative estimate on the significance vs luminosity not because of the signal but for the higher cross-section (via VBF) of $W^+W^-Z$ compared to $W^+W^-H$ \cite{Costantini:2020stv}. The pair production $h_P^+h_P^-$ has been considered background-free. The pure charged triplet $h_P^\pm$ has a single decay channel, namely $h_P^+\to a_P (W^+)^*$. Whereas the pseudoscalar is undetectable, the process will give rise to displaced off-shell $W$ boons: there is no SM process that have this particular final state.  This also gives rise to displaced leptons/jets plus missing energy in the final-state.

In Figure~\ref{fig:zhtsig}(b) we plot the reach of the triplet VEV $v_T$ as a function of $\mathcal{L}_{5\sigma}$. This is the luminosity required at a 14 TeV muon collider for the discovery of $h_T$ produced via VBF. The estimation of the background has been already explained. The black dot-dashed line is obtained from our scanned points that satisfy the Planck constraint on the DM relic density.

\section{Conclusions}\label{sec:concl}
Here we studied an extension of the SM with a complex hyperchargeless triplet scalar. The triplet extensions are well motivated from the viewpoint of the enhanced vacuum stability \cite{Khan:2016sxm,Jangid:2020qgo}. On top of that a complex triplet extension of the Standard Model can provide a natural dark matter candidate without any discrete symmetry because the purity of the triplet acts as an odd number in a $Z_2$ symmetry.
 The scalar spectrum has 4 additional scalars, 2 neutral and 2 charged. In the cTSM the massive pseudoscalar is a pure triplet and its pureness makes it a DM candidate as it fails to have any cubic interaction vertex with the fermions as well as the gauge bosons. In the $Z_3$ symmetric limit the pure charged Higgs boson and pure pseudoscalar become degenerate in mass at the tree-level. The relic density constraint from Planck measurements is satisfied if $m_{a_P}\sim1.35-1.60$ TeV for the scanned data points which also satisfy the direct dark matter constraints as well as the LHC Higgs boson data.
 
 Apart from the pseudoscalar, the spectrum consist of two charged Higgs bosons almost degenerate in mass: $h_P^\pm$ and $h_T^\pm$. The former is a pure triplet whereas the later is a state with a small mixing with the doublet. The purity conservation prohibits any 2-body decays $h_P^\pm$, making the collider phenomenology of the cTSM quite interesting. The $h_T^\pm$ has prompt decay into the non-standard mode of $ZW^\pm$, which is a signature of custodial breaking \cite{Espinosa:1991wt,Godbole:1994np} and can be differentiated from other non-standard mode like $h^\pm\to a\,W^\pm$ in the case of NMSSM charged Higgs boson \cite{Bandyopadhyay:2015dio}, or $h^\pm\to N\,e^\pm$ in case of Type-X with right-handed neutrino \cite{Bandyopadhyay:2019xfb}. In the case of superysmmetric extensions with triplets, physical charged Higgs bosons are always mixed with the doublet ones and so are the pseudoscalars. This lack of pureness of the triplet states implies that the possibility of a pseudoscalar dark matter ceases to exist \cite{Bandyopadhyay:2014vma,Bandyopadhyay:2015ifm}. 
  
In the cTSM the most interesting charged Higgs boson is the pure triplet one, i.e. $h_P^\pm$ which decays to $a_P (W^\pm)^*$. This can give rise to displaced charged leptons/jets plus missing energy \cite{Jangid:2020qgo}. We give an estimate of that at a multi-TeV muon collider along with many other production channels. Such lightest charged Higgs has a lifetime $\tau_{h_P^\pm}=\mathcal{O}(1)$ m, in the range of proposed experiments for testing long-lived particles like MATHUSLA \cite{Lubatti:2019vkf,Jangid:2020qgo}. The triplet playing a role in EWSB  can be estimated by probing the $v_T$ (VEV of the triplet)  and the corresponding required luminosities are also listed. The LHC and FCC would look for the cubic and quartic Higgs couplings and the triplet contamination can also be constrained.

The $Y=0$ triplet nature makes these excitation to hard to be produced at present hadron colliders. Thus the next elusive Higgs may be quite natural. Looking for the mentioned channels at the LHC with 14 and 100 TeV centre of mass energy along with the futuristic multi-TeV muon collider can provide us with some surprises.

\section*{Acknowledgements} PB wants to thank Anomalies 2019 and Anomalies 2020 for the reason of this collaboration. PB acknowledges SERB CORE Grant CRG/2018/004971 for the support. PB also thanks Aleesha KT for the help in PYTHIA8 and Dr. Shubho Roy for some useful discussions. The work of AC is supported by INFN research grant n. 20286/2018.

\appendix

\section{Renormalization Group Equations}\label{sec:rge}
We list the renormalization group equation for the dimensionless parameter of the cTSM, computed at one-loop \cite{Du:2018eaw}. We have
\bea
16\pi^2\frac{d}{dt}g_1&=&\frac{41}{10}g_1^3\\
16\pi^2\frac{d}{dt}g_2&=&-\frac{5}{2}g_2^3\\
16\pi^2\frac{d}{dt}g_3&=&-7g_3^3\\
16\pi^2\frac{d}{dt}y_t&=&y_t\Big(\frac{9}{2}y_t^2-\frac{17}{20}g_1^2-\frac{9}{4}g_2^2-8g_3^2\Big)\\
16\pi^2\frac{d}{dt}\lambda_H&=&\lambda_H\Big(12\lambda_H+12y_t-\frac{9}{5}g_1^2-9g_2^2\Big)+\frac{3}{2}\lambda_{HT}^2-12y_t^4+\frac{27}{100}g_1^4+\frac{9}{10}g_1^2g_2^2+\frac{9}{4}g_2^4\\
16\pi^2\frac{d}{dt}\lambda_{HT}&=&\lambda_{HT}\Big(2\lambda_{HT}+6\lambda_H+8\lambda_T+6\lambda_{T'}+6y_t^2-\frac{9}{10}g_1^2-\frac{33}{2}g_2^2\Big)+12g_2^4\\
16\pi^2\frac{d}{dt}\lambda_T&=&\lambda_T\Big(14\lambda_T+12\lambda_{T'}-24g_2^2\Big)+\lambda_{HT}^2+30g_2^4+3\lambda_{T'}^2\\
16\pi^2\frac{d}{dt}\lambda_{T'}&=&\lambda_{T'}\Big(9\lambda_{T'}+12\lambda_T-24g_2^2\Big)-12g_2^4
\eea

\bibliography{obscurum_refs}

\end{document}